\documentclass[nofootinbib,a4paper]{revtex4}
\usepackage[paperwidth=210mm,paperheight=297mm,centering,hmargin=2cm,vmargin=2.5cm]{geometry}
\usepackage{amsmath,amssymb,amsfonts}
\usepackage{graphicx}

\usepackage{tikz}
\usepackage{xfp} % for fpeval -> floating point expression
\usepackage{varwidth} % for controlling maximum width of boxes in nodes

\usepackage[noEnd = false]{algpseudocodex}

\usepackage[normalem]{ulem}

\usepackage{tabulary} %% added
\usepackage{booktabs}

\usepackage{comment}

\usepackage{orcidlink}

\usepackage{xspace}

\usepackage{mathrsfs}
\DeclareSymbolFontAlphabet{\mathrsfs}{rsfs}
\newcommand{\scri}{\mathrsfs{I}}
\newcommand{\scrip}{$\scri^+$}

\newcommand{\Kc}{K_{\textrm{CMC}}}
\newcommand{\aconf}{\Omega}
\newcommand{\rscri}{r_\scri}
\newcommand{\aaa}{\kappa}
\newcommand{\minusaaa}{-\aaa}
\newcommand{\aaasquared}{\aaa^2}
\newcommand{\amplitude}{A}
\newcommand{\C}{C}
\newcommand{\shorthand}{h}

\newcommand{\eref}[1]{(\ref{#1})}
\newcommand{\sref}[1]{section~\ref{#1}}

\newcommand{\ssref}[1]{subsection~\ref{#1}}
\newcommand{\aref}[1]{appendix~\ref{#1}}
\newcommand{\fref}[1]{figure~\ref{#1}}

\begin{document}

\title{Hyperboloidal approach for linear and non-linear wave equations in FLRW spacetimes}

\author{Flavio Rossetti\,\orcidlink{0000-0001-5500-2458}$^{1, 2}$}
\thanks{\textit{E-mail address}: \texttt{flavio.rossetti@gssi.it}}
\author{Alex Vañó-Viñuales\,\orcidlink{0000-0002-8589-006X}$^{3}$}
\address{$^1$Gran Sasso Science Institute, Viale Francesco Crispi 7, L'Aquila (AQ), 67100, Italy}
\address{$^2$CAMGSD, Departamento de Matemática, Instituto Superior Técnico IST, Universidade de Lisboa UL, Avenida Rovisco Pais 1, 1049-001 Lisboa, Portugal}
\address{$^3$Centro de Astrof\'{\i}sica e Gravita\c c\~ao - CENTRA, Departamento de F\'isica, Instituto Superior T\'ecnico IST, Universidade de Lisboa UL, Avenida Rovisco Pais 1, 1049-001 Lisboa, Portugal}%\ead{flavio.rossetti@tecnico.ulisboa.pt} % update Flavio's email?
%\author*[1]{\fnm{Flavio} \sur{Rossetti\,\orcidlink{0000-0001-5500-2458}}}\email{flavio.rossetti@tecnico.ulisboa.pt} \affil*[1]{\orgdiv{CAMGSD, Departamento de Matem\'atica}, \orgname{ Instituto Superior T\'ecnico IST, Universidade de Lisboa UL}, \orgaddress{\street{Avenida Rovisco Pais 1}, \postcode{1049}, \city{Lisboa}, \country{Portugal}}}  \author[1]{\fnm{Alex} \sur{Va\~n\'o-Vi\~nuales\,\orcidlink{0000-0002-8589-006X}}}\email{alex.vano.vinuales@tecnico.ulisboa.pt}    \affil[1]{\orgdiv{CENTRA, Departamento de F\'isica}, \orgname{ Instituto Superior T\'ecnico IST, Universidade de Lisboa UL}, \orgaddress{\street{Avenida Rovisco Pais 1}, \postcode{1049}, \city{Lisboa}, \country{Portugal}}}

\begin{abstract}
%\abstract{

In this numerical work, we deal with two distinct problems concerning the propagation of waves in cosmological backgrounds. In both cases, we employ a spacetime foliation given in terms of compactified hyperboloidal slices. These slices intersect \scrip, so our method is well-suited to study the long-time behaviour of waves. Moreover, our construction is adapted to the presence of the time--dependent scale factor that describes the underlying spacetime expansion.

 First, we investigate decay rates for solutions to the linear wave equation in a large class of expanding FLRW spacetimes, whose non--compact spatial sections have either zero or negative curvature.   By means of a hyperboloidal foliation, we provide new numerical evidence for the sharpness of decay--in--time estimates for linear waves propagating in such spacetimes.
% is more robust than the method of truncated Cauchy slices we exploited in \cite{Rossetti:2023igb}, and allows to retrieve the same decay results.

Then, in the spatially-flat case, we present numerical results in support of small data global existence of solutions to semi-linear wave equations in FLRW spacetimes having a decelerated expansion, provided that a generalized null condition holds. In absence of this null condition and in the specific case of $ \square_g \phi = (\partial_t \phi)^2 $ (Fritz John's choice), the results we obtain suggest that, when the spacetime expansion is sufficiently slow, solutions diverge in finite time for every choice of initial data. 

%}
\end{abstract}

%\maketitle

\maketitle

%%%%%%%%%%%%%%%%%%%%%%%%%%%%%%%%%%%%%%%%%%%%%%%%%%%%%%%%%%
\section{Introduction}

Due to the hyperbolic nature of Einstein's equations, the simplest model apt to investigate stability and instability of spacetimes is given by the wave equation on a fixed background. While decay results for wave solutions (and their derivatives) provide crucial insights in this direction, non-linear terms can be further added to study a more realistic self-interacting framework. 
 In this article we report numerical experiments based on non-traditional spacetime slices, aimed to study solutions to the linear and non-linear wave equation for background choices of cosmological interest. The limitations of numerical studies is that results can support mathematical estimates or small data global existence proofs, but cannot be as conclusive as mathematically-obtained evidence. Still, they can be invaluable in providing physical intuition and/or guiding analytical research.

In the following, given the Friedmann--Lema{\^i}tre--Robertson--Walker (FLRW) metric
\[
g = -d \tilde t^2 + a(\tilde t)^2 d\Sigma_3^{\, 2}%\left ( d \tilde r^2 + \psi(\tilde r)^2 d\sigma^2 \right),
\]
with scale factor $a(\tilde t) > 0$ describing a decelerated expansion
%, with $d\sigma^2$ standard metric on $S^2$ and where either $\psi(\tilde r) = \tilde r$ (spatially--flat case) or $\psi(\tilde r)=\sinh( \tilde r)$ (spatially--hyperbolic case)
 and a three-dimensional Riemannian metric $d\Sigma_3^{\, 2}$, we employ numerical methods to analyse spherically symmetric\footnote{Assuming spherical symmetry is a natural choice due to the homogeneity and isotropy of the underlying metric.} solutions to the linear wave equation
\begin{equation} \label{wave_eqn}
\square_g \phi = 0.
\end{equation}
We deal with non--compact spatial sections given either by $\mathbb{R}^3$ or by the hyperbolic space $\mathbb{H}^3$.

In the case of flat spatial sections ($d\Sigma_3^{\, 2}=d \tilde r^2 + \tilde r^2 d\sigma^2$, with $d\sigma^2$  the standard metric on $S^2$), we also investigate spherically symmetric solutions to the non-linear wave equation
\[
\square_g \phi = {\tilde{\mathcal{N}}}^{\alpha \beta} \partial_{\alpha}\phi \partial_{\beta}\phi,
\]
where 
\begin{align*}
{\tilde{\mathcal{N}}}^{\tilde t\tilde t} &= c_1 a(\tilde t)^{\beta_1}, \\
{\tilde{\mathcal{N}}}^{\tilde r\tilde r} &= c_2 a(\tilde t)^{\beta_2}, \\
{\tilde{\mathcal{N}}}^{\tilde r\tilde t} = {\tilde{\mathcal{N}}}^{\tilde t\tilde r} &= \frac{c_3}{2} a(\tilde t)^{\beta_3},
\end{align*}
with $c_1, c_2, c_3 \in \mathbb{R}$ and $\beta_1, \beta_2, \beta_3 \in \mathbb{R}$. Here we highlight that the $(\tilde t, \tilde r)$ chart is different  from the compactified coordinate system $(t, r)$ used in the numerical simulations, see already section \ref{subsec:flatconformal}.  In constrast, we  never rescale the metric tensor.
The time-dependent weights for the non-linear terms are inspired by \cite{CostaFranzenOliver} and the specific choices we made for  $c_i,$ $\beta_i$,  $i=1, 2, 3$ are discussed further in section \ref{section:nonlinearwave}. 

These equations can be seen as a family of geometric wave equations on Lorentzian manifolds that describe homogeneous and isotropic expanding spacetimes. 
The above  equations can also be regarded as a class of wave equations in $\mathbb{R}^{1+3}$ with time--dependent damping terms and, more generally, time--dependent coefficients, see e.g. \cite[Section 6]{NatarioRossetti} and \cite{Wirth}. As seen in \cite{NatarioRossetti, CostaNatarioOliveira}, waves propagating in expanding FLRW spacetimes are affected by  dispersion and cosmological redshift. In the spatially--flat case  with scale factor $a(\tilde t) = \tilde t^p$, the former determines the $L^{\infty}$ decay in the scenario of slow expansions. For faster expansion rates, on the other hand, the leading--order contribution to the decay is due to cosmological redshift.\footnote{Notice that, when   $d\Sigma_3^{\, 2}$ is the standard metric on $\mathbb{R}^3$ and $a(\tilde t)=\tilde t^p$, the non--expanding case $p = 0$ corresponds to Minkowski spacetime and the decay is entirely due to dispersion.} In \cite{Rossetti:2023igb}, these two contributions could be isolated in different spacetime regions.

\textbf{The main results of this work are the following:}
\begin{enumerate}
\item In the linear setting, we retrieve the decay estimates of \cite{Rossetti:2023igb} by exploiting hyperboloidal numerical schemes, which are advantageous with respect to traditional methods. Our results provide a further numerical confirmation of the sharpness of these rates of decay.
\item In the non--linear setting, we find evidence for small data global existence results in the context of FLRW spacetimes having decelerated expansion. 
\end{enumerate}

With respect to the latter, our numerical results expand the rigorous work \cite{CostaFranzenOliver}, where small data global existence was investigated in the spatially--flat, accelerating regime (i.e. $a(\tilde t)=\tilde t^p$, with $p > 1$). In the decelerating case ($a(\tilde t)=\tilde t^p$, with $0 < p < 1$)  that we focus on, our new numerical results suggest that small data global existence holds for non--trivial, smooth solutions to 
\begin{equation} \label{gennullcondition}
\square_g \phi = (\partial_{\tilde t} \phi)^2 - \frac{1}{a(\tilde t)^2} (\partial_{\tilde r} \phi)^2,
\end{equation}
i.e. a generalized null condition seems to hold.
On the contrary, our numerical experiments show that solutions to
\begin{equation} \label{fritzjohn}
\square_g \phi = (\partial_{\tilde t} \phi)^2,
\end{equation}
diverge in finite time for every choice of compactly supported initial data if the expansion is sufficiently slow ($0 < p < \epsilon$, for some $\epsilon > 0$ small). This behaviour is consistent with the $p=0$ (Minkowski) case. Larger values of $p$ (i.e. a faster expansion) seem to be compatible with small data global existence, specifically in the interval $p \ge \frac12$. 
Whether such a value is the threshold of the two behaviours above is still an open question.

Regarding our numerical implementation, it uses the hyperboloidal approach. Its innovative merit is that instead of evolving on traditional Cauchy slices, the background is expressed in terms of hyperboloidal slices -- spacelike slices that reach future null infinity (\scrip), the collection of end-points of future-directed null geodesics.  See the conformal diagrams on \fref{fig:time_dep_height} for a depiction.
 In spacetimes where future null infinity is an ingoing null hypersurface (valid e.g.~for $0\le p< 1$ in the flat FLRW case), no physical information enters the domain from beyond \scrip\xspace and signals just leave the computational domain. The advantage of this is that if \scrip\xspace is located at the outer boundary of the computational domain, no physical boundary conditions are to be imposed. This solves the generally difficult issue of spurious contributions originating from the boundary due to imperfect boundary conditions. Computationally, this setup also gives more efficient results, as an infinite physical distance is covered by finite coordinate values and physically sparse spatial grids are enough to properly resolve the propagating signals at all times. 

From a numerical relativity perspective, hyperboloidal methods are the current state-of-the-art of solving the Teukolsky equation in perturbation theory \cite{Zenginoglu:2011jz,Harms:2013ib,PanossoMacedo:2019npm}. Besides useful toy models e.g.~\cite{Hilditch:2016xzh,Gasperin:2019rjg}, evolution of the non-linear Einstein equations on hyperboloidal slices is a long-awaited milestone towards which steady progress is taking place \cite{Zenginoglu:2008pw,Vano-Vinuales:2014koa,Peterson:2024bxk}.

Regarding mathematical applications, hyperboloidal methods are a useful tool to study the smoothness of null infinity \cite{Andersson:1992yk}. A further area of application concerns global existence results for hyperbolic equations in curved spacetimes in those cases where the symmetry group of the equation under study is not necessarily given by the Poincaré group \cite{FlochMa}.
In non-linear settings, hyperboloidal coordinates have been exploited to extend (weak) solutions to semi-linear wave equations beyond the occurrence of singularities, see for instance the recent mathematical results in \cite{BiernatDonningerSchorkhuber, DonningerOstermann}. 

 Our hyperboloidal setup is described in \sref{section:hypapproach}, and the actual implemented wave equations for the considered cases are given in \aref{eqsappendix}. The algorithm to recover the expected decay rates is sketched in \sref{section:alg_explained}, with a detailed explanation in \aref{appendixAlgorithm}. Our results for the linear wave equation are presented in \sref{linearresults} and those for the non-linear experiments in \sref{section:nonlinearwave}. Finally we provide some concluding remarks.

%%%%%%%%%%%%%%%%%%%%%%%%%%%%%%%%%%%%%%%%%%%%%%%%%%%%%%%%%%
\section{Hyperboloidal approach} \label{section:hypapproach}

The basic ingredients to translate from a traditional Cauchy evolution to a hyperboloidal one reaching \scrip\xspace are a transformation in the time coordinate, so that its level sets intersect future null infinity instead of spatial infinity, and the compactification of the spatial coordinates, putting \scrip\xspace at a finite coordinate distance. %From now onwards, $\tilde t$ will denote the usual time coordinate and $t$ the hyperboloidal time. They are related via the relation
Taking $\tilde T$ to denote our time coordinate and $T$ to be the hyperboloidal time, they are related via
\begin{equation} \label{genericttrafo}
\tilde T = T  +h(\tilde x^i),%\tilde t = t  +h(\tilde x^i),
\end{equation}
where the quantity $h$ is called the height function, and  is to be chosen so that the hyperboloidal time $T$ asymptotically (for infinite radial coordinate) behaves as the retarded time $\tilde u$ for a slice reaching \scrip (see also \cite{Zenginoglu:2024} for a more precise definition). If it were to reach $\scri^-$, $T$ would asymptotically  behave like the advanced time $\tilde v$. 
The transformation in the radial-like coordinate is given by a time-independent compress (or compactification) factor that allows to reach null infinity with finite coordinate distance. %$\aconf(t,r)$
%quantities with tilde correspond to uncompactified, traditional coordinates
 Given the time-dependent character of the FLRW metrics, it is useful to consider $h(t,\tilde x^i)$ (or $h(t,x^i)$) and a compactification including time-dependence. Nevertheless, this makes the resulting expressions very complicated and it is not clear how to impose the introduced freedom in a useful way. We have developed a prescription for a time-dependent function and time-independent compactification (see \ssref{subsection:flatphysical}) that worked successfully. Further generalising this setup may come in handy in future work.
%In principle, given the time-dependent character of the FLRW metrics considered, we could also consider $h(t,\tilde x^i)$ (or $h(t,x^i)$) and a compactification including time-dependence. Nevertheless, this makes the resulting expressions very complicated and it is not clear how to impose the introduced freedom in a useful way. This generalization may however come in handy in future work. 

\subsection{Flat case using conformal time} \label{subsec:flatconformal} %Wave equation on c %Compactified hyperboloidal slice for the 

The line element of a FLRW spacetime with flat spatial sections and time-dependent scale factor $a$ is given by the following expression, where $\tilde t$ denotes the usual time coordinate, $\tilde r$ the unrescaled radial coordinate, and $d\sigma^2=d\theta^2+\sin^2\theta d\varphi^2$ is the standard metric on $S^2$:
\begin{equation} \label{FLRWflatmetric}
ds^2 = -d\tilde t^2+a^2(\tilde t) \left(d\tilde r^2 + \tilde r^2d\sigma^2\right). 
\end{equation}
It is however more convenient to work with the conformal time $d\tilde\tau=d\tilde t/a$:
\begin{equation}\label{dsflatconf}
ds^2 = a^2(\tilde\tau) \left(-d\tilde\tau^2+d\tilde r^2 + \tilde r^2d\sigma^2\right). 
\end{equation}
The scale factor in the flat case takes the form
\begin{equation}\label{scalefacflat}
a(\tilde t) = \tilde t^p \quad \Leftrightarrow\quad a(\tilde\tau) = \left[(1-p)\tilde\tau\right]^{\frac{p}{1-p}}. 
\end{equation}
When changing to the hyperboloidal slicing, the transformation in the time coordinate \eref{genericttrafo} will be applied on the conformal time as 
\begin{equation} \label{tautrafo}
\tilde \tau = \tau+h(\tilde r).
\end{equation}
The compactification of the radial coordinate is written in the following form in our language, with $\aconf$ the compactification factor:
\begin{equation} \label{compactflat}
\tilde r = \frac{r}{\aconf(r)}. 
\end{equation}
After applying transformations \eref{tautrafo} and \eref{compactflat} to \eref{dsflatconf}, the line element becomes
\begin{equation}\label{fsthyp}
ds^2= a^2(\tau+h) \left[-d\tau^2-2\partial_{\tilde r}h\,\frac{(\aconf-r\,\aconf')}{\aconf^2}d\tau\,dr+\left[1-\left(\partial_{\tilde r}h\right)^2\right]\frac{(\aconf-r\,\aconf')^2}{\aconf^4}d r^2 + \frac{r^2}{\aconf^2} d\sigma^2\right] ,
\end{equation}
where $h=h(\tilde r)$ and $\aconf=\aconf(r)$, see already \eqref{hfuncflat} and \eqref{confflat} for their expressions. 

Defining the parameter $\aaa$, which is related to the constant value of the trace of the physical extrinsic curvature as
\begin{equation} \label{kappaK}
\aaa = \frac{-3}{\Kc}, 
\end{equation}
we choose the same height function exploited in Minkowski spacetime to transform to constant-mean-curvature (CMC) slices \cite{Malec:2009hg}, as used in e.g~\cite{Zenginoglu:2007jw}:
\begin{equation}\label{hfuncflat}
%h(\tilde r) = \sqrt{\left(\frac{3}{\Kc}\right)^2+\tilde r^2}+\frac{3}{\Kc} \quad \Leftrightarrow\quad h(r) = \sqrt{\left(\frac{3}{\Kc}\right)^2+\left(\frac{r}{\aconf}\right)^2}+\frac{3}{\Kc}
h(\tilde r) = \sqrt{\aaasquared+\tilde r^2}\minusaaa \quad \Leftrightarrow\quad h(r) = \sqrt{\aaasquared+\left(\frac{r}{\aconf}\right)^2} \minusaaa.
\end{equation}
The last term in the expression of the height function makes it zero at the origin of the spatial coordinates. Its presence has no impact on the metric, as only the spatial derivative of $h$, the boost, appears there, but it induces a time translation in the scale factor. Still, this has no relevant impact in the  results. 

Our numerical experiments were mostly performed using the above height function. A time--dependent height function, which can be employed for a more general class of equations with time--dependent terms, is introduced in section \ref{subsection:flatphysical} and was successfully tested for the spacetimes taken in consideration. 

It is useful, especially when verifying the spacelike character of the foliation, to write the hyperboloidal time in terms of a double null coordinate system  $(\tilde u, \tilde v)$, with $\tilde u = \tilde \tau -\tilde r$, $\tilde v = \tilde \tau  + \tilde r$  and the height function as specified in \eref{hfuncflat}:
\[
\tau = \tilde \tau -h(\tilde u, \tilde v) = \frac{\tilde u + \tilde v}{2} +  \kappa - \sqrt{\kappa^2 + \frac{(\tilde u - \tilde v)^2}{4}} .
\]
It is immediate to verify that $X = \frac{\partial \tau}{\partial \tilde u} \partial_{\tilde u} + \frac{\partial \tau}{\partial \tilde v} \partial_{\tilde v} $, which is the vector field orthogonal to the level sets $\{\tau = \text{const.}\}$, is timelike.

%show how the above gives the correct asymptotic behaviour $\tau\to\tilde u$ -- review Flavio's calculations 
%{\fr Added by Flavio, the height function used here does *not* contain the additional term $3/K_{\text{CMC}}$:}
%In terms of the $\tilde u$, $\tilde v$ null coordinates, we have
%\[
%\tau = \tilde \tau -h(\tilde u, \tilde v) = \frac{\tilde u + \tilde v}{2} + \frac{3 \left  ( 1 + \left  (\frac{6}{\Kc (\tilde v - \tilde u)} + \sqrt{1 + \frac{36}{\Kc^2(\tilde v - \tilde u)^2}} \right ) ^2 \right ) }{\Kc \left ( 1 - \left  ( \frac{6}{\Kc (\tilde v - \tilde u)} + \sqrt{1 + \frac{36}{\Kc^2(\tilde v - \tilde u)^2}} \right ) ^2 \right ) }.
%\]
%The timelike vector field orthogonal to the curves $\{\tau = \text{const.}\}$ is $X = \frac{\partial \tau}{\partial \tilde u} \partial_{\tilde u} + \frac{\partial \tau}{\partial \tilde v} \partial_{\tilde v} $, which is asymptotically null in the limit $\tilde v \to +\infty$. {\fr Flavio: Not quite true: I forgot the scale factor!}

The compactification factor we choose, compatible with the height function \eref{hfuncflat}, is
\begin{equation} \label{confflat}
\aconf(r)= \frac{\rscri^2-r^2}{2\ \aaa\ \rscri} %-\Kc\frac{\rscri^2-r^2}{6\ \rscri},
\end{equation}
where we set $\rscri=1$ without loss of generality. 
After substitution of all the above coordinate transformations and values, the final expression for the linear wave equation \eref{wave_eqn} is \eref{eflat}, included in \aref{eqsappendix}.

The amplitude of the wave function decays asymptotically so that it becomes zero at \scrip. If we evolve the unrescaled scalar field we indeed observe that the field becomes zero as it reaches the boundary of the domain. It is also possible to evolve a rescaling of $\phi$. For instance, $\bar\phi = \phi/\aconf$ is such that for Minkowski the scalar field is $O(1)$ at \scrip. However, for non-zero small $p$, the scalar field shows steep gradients at \scrip\xspace and the data do not look smooth near future null infinity. %, which easily induce loss of convergence. 
Rescaling the field as $\hat\phi = \phi/\aconf^{\frac{1}{1-p}}$ provides non-vanishing finite amplitudes at \scrip\xspace for every value of $p\in[0,1)$. However, in our experiments we found that the simulations blow up starting at \scrip\xspace in finite time for any value of $p$, and this time is earlier for larger $p$.  While studying this effect in detail is beyond the scope of this work, we experimented with various aspects of the setup, such as increasing the value of $\aaa$, which slows down the simulations and blow--up happens later. From a convergence point of view, different choices perform differently at the numerical level -- see the observations in \sref{sec:convergence}.

\subsection{Flat case using physical time with time-dependent height function} \label{subsection:flatphysical}

%{\avv as the other two setups use conformal time, would it make sense to put them together and start with this one? Also we need to do some numerical tests that allow us to recover the expected decay.}

\begin{figure}[h]
\begin{center}
\includegraphics[width=0.32\linewidth]{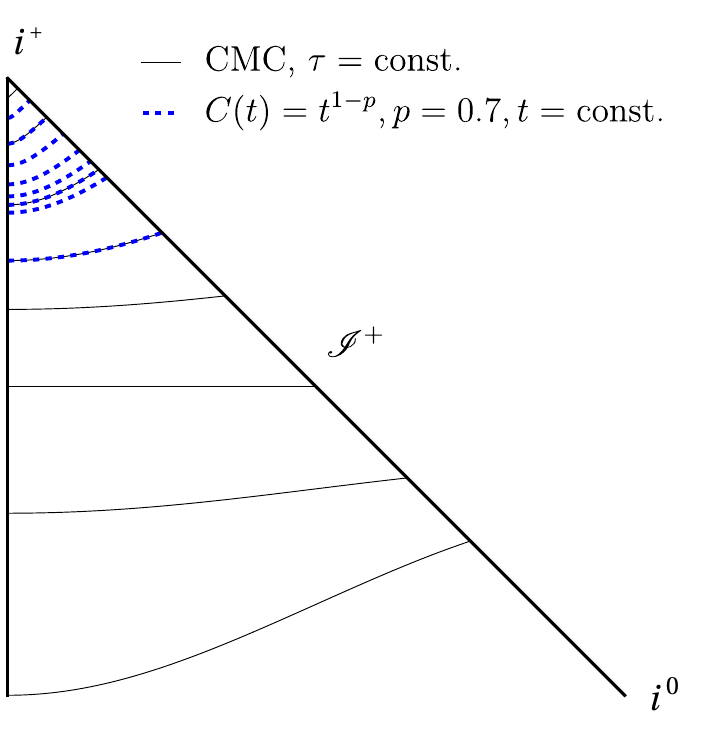}
\includegraphics[width=0.32\linewidth]{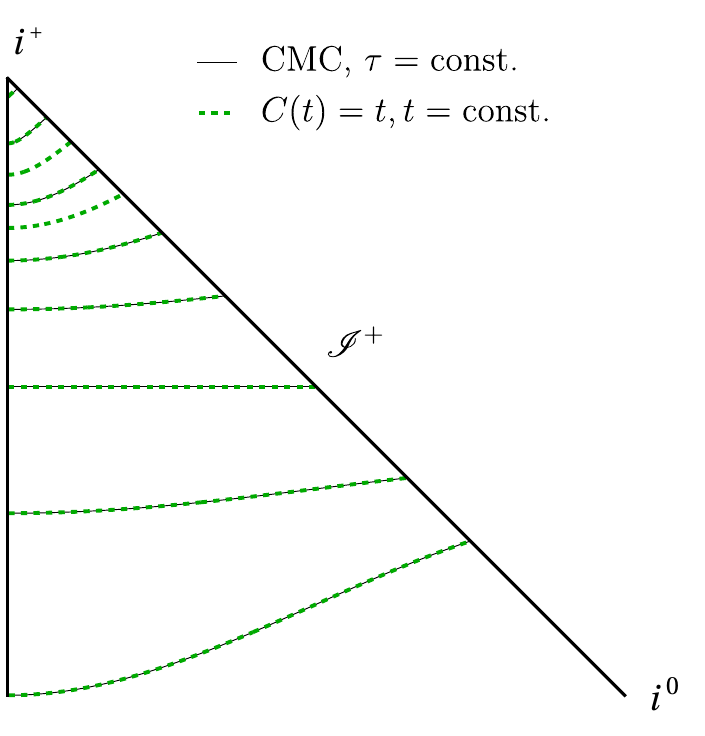}
\includegraphics[width=0.32\linewidth]{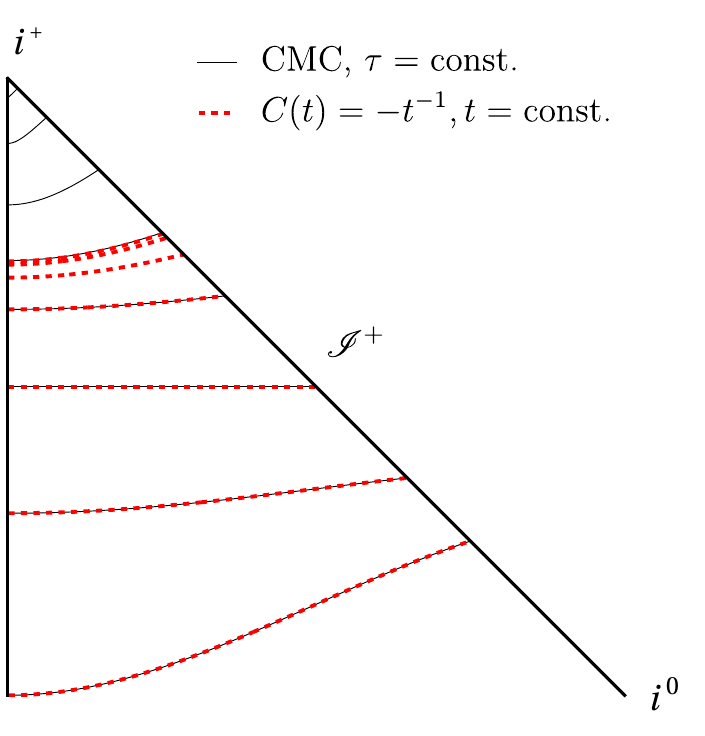}
\caption{Conformal Carter-Penrose diagrams depicting the hyperboloidal slices considered in the spatially flat case when the height function is time--dependent. In the horizontal axis $R = (V-U)/2$ is plotted and $T = (V+U)/2$ in the vertical one, with the common choice $U=\arctan(\tilde u)$, $V=\arctan(\tilde v)$ and $\tilde u = \tilde \tau-\tilde r$, $\tilde v = \tilde \tau+\tilde r$. For more details on the construction of Penrose diagrams for hyperboloidal slices see e.g.~\cite{Vano-Vinuales:2023pum}. The black lines correspond to level sets of the hyperboloidal time $\tau$   given by \eref{tautrafo} with the CMC height function \eref{hfuncflat}, with $\aaa=1$ used in all cases. The dashed lines correspond to foliations given by hyperboloidal time $t$ following the transformation \eref{ttrafo} substituted into $\tilde\tau = \tilde t^{1-p}/(1-p)$ and using the time-dependent height function \eref{tdepheightfunc}. Each color corresponds to a different choice of $C(t)$ as indicated in the legend.} \label{fig:time_dep_height} 
\end{center}
\end{figure}

Here we briefly describe a possible way to define a time--dependent height function. We keep our expressions as simple as possible by assuming that the compactification factor does not depend on time. By using \eref{genericttrafo} as applied on the physical time $\tilde t$
\begin{equation} \label{ttrafo}
\tilde t = t+ h(t, \tilde r),
\end{equation}
imposing that (see also the line element \eref{ansatzflatphysical} we compare our metric with):
\begin{equation}
\frac{\gamma_{rr}}{\frac{a^2}{\aconf^2}} = \frac{ a^2 \left(\aconf-r \aconf'\right)^2-\aconf^4 \left(h'\right)^2}{a^2\aconf^2} \equiv 1 = \frac{\gamma_{\theta\theta}}{\frac{a^2}{\aconf^2}}
\end{equation}
and substituting  the time-independent \eref{confflat} as expression for the compactification factor, we obtain a differential equation for $h$, expressed in terms of the compactified radial coordinate $r$. After choosing a suitable sign:
\[
h'=\partial_{r}h(t, r) = \frac{4 \kappa r}{(1-r^2)^2} \left( t + h(t, r) \right)^p . 
\]
This yields the following expression for the height function:
\begin{equation}\label{tdepheightfunc}
h = \left[(p-1)\left(\frac{-2\aaa}{1-r^2}-\C(t)\right)\right]^{\frac{1}{1-p}} -t,
\end{equation}
where $\C(t)$ is a time-dependent integration constant. %We choose it to be $\C(t)=t^{1-p}$, since this expression is convenient when switching between physical and conformal time and allows to have real--valued quantities. A drawback of this choice, on the other hand, is that the larger the $p$ {\fr the slower these slices} reach higher values of the retarded time coordinate {\fr $\tilde u$ when comparing to the $(\tilde u, \tilde v)$ system}. A depiction of this foliation for $p=0.3$, in comparison with the usual Minkowski foliation is in Fig. \ref{fig:time_dep_height}.
We consider three different choices: $\C(t)=t^{1-p}$, $\C(t)=t$ and $\C(t)=-t^{-1}$, whose level sets are depicted in conformal-diagram form in \fref{fig:time_dep_height}. For the time intervals under investigation, the above height function is always real--valued.  These choices for $C(t)$ are mainly motivated by mathematical convenience. The first option for $C(t)$ is convenient when switching between physical and conformal time, but a relative drawback is that the larger $p$ the slower these slices reach higher values of the retarded time coordinate $\tilde u$ when comparing to the $(\tilde u, \tilde v)$ system. This is the only $C(t)$ choice that depends on the value of $p$, and in the left diagram of \fref{fig:time_dep_height} the value $p=0.7$ has been set. The second option for $C(t)$ is able to cover the whole domain with its foliation (middle figure), while in the third case the slices accumulate at a maximum value of $t$ (see right figure). The slices obtained for these time-dependent height functions coincide (at least in some regions) with CMC ones, which are shown by the solid black lines. 
%$\C(t)=-t^{-1}$. 
%{\avv from the notes I gather that this choice of $\C(t)$ (page 7 in Flavio's notes) is based on a beneficial property on $\tilde\alpha$, but I forgot exactly why}: it has to be time-dependent and between -2 and 0.

The linear wave equation expressed in terms of these height functions is given respectively by \eref{eflavio1}, \eref{eflavio2} and \eref{eflavio3} in the appendix. 

%{\fr Time-dependent height function:} I retrieved the same decay rates. Since the foliation is different, different parameters might be needed. For instance, for p = 0.8, I used $Kcmc=-18$ (!), end\_time $=$ 500 , every\_t\_step = 5000 (this already takes 15 mins on my laptop) and got decay $\alpha \sim 0.68$ for $\log \tilde t \in [-2, -0.5]$.

\subsection{Hyperbolic case using conformal time}

The line element for the spatially hyperbolic slices is given by
\begin{eqnarray}
ds^2 &=& -d\tilde t^2+a^2(\tilde t) \left(d\tilde \rho^2 + \sinh^2\tilde \rho\ d\sigma^2\right)\\ &=&  a^2(\tilde\tau) \left(-d\tilde\tau^2+d\tilde \rho^2 + \sinh^2\tilde \rho\ d\sigma^2\right). 
\end{eqnarray}
The scale factor in terms of the conformal time is 
\begin{equation} \label{scalefachyperb}
a(\tilde\tau) = \left[\sinh \left(\frac{3w+1}{2}\tilde\tau \right) \right]^{\frac{2}{3w+1}},
\end{equation}
where $0 \le w < 1$. Strictly speaking, the Friedmann equations imply that the constraint $w > -\frac13$ is sufficient to have a decelerated universe. Here, we further restrict our analysis to non--negative values of $w$, i.e. to a real--valued speed of sound. We define the compactified radial coordinate $\rho$ by
\begin{equation} \label{compacthyperb}
\sinh\tilde \rho = \frac{\sinh\rho}{\aconf(\rho)}.
\end{equation}
In terms of the above and of \eref{tautrafo}, the line element becomes:
\begin{equation}\label{hyperhyp}
ds^2= a^2(\tau+h) \left[-d\tau^2-2\partial_{\tilde \rho}h\,\frac{(\aconf\cosh\rho-\aconf'\sinh\rho)}{\aconf^2\sqrt{1+\frac{\sinh^2\rho}{\aconf^2}}}d\tau\,d\rho+\left[1-\left(\partial_{\tilde \rho}h\right)^2\right]\frac{(\aconf\cosh\rho-\aconf'\sinh\rho)^2}{\aconf^2\left(\aconf^2+\sinh^2\rho\right)}d \rho^2 + \frac{\sinh^2\rho}{\aconf^2} d\sigma^2\right] ,
\end{equation}
%$a = a(\tau+h(\sinh^{-1}\left[\frac{\sinh\rho}{\aconf}\right]))$ and 
where $h=h(\tilde \rho)$ and $\aconf=\aconf(\rho)$, see already \eqref{hfunchyperb} and \eref{confhyperb} for their expressions.

Analogously to section \ref{subsec:flatconformal}, we define the height function as:
\begin{equation}\label{hfunchyperb}
%h(\tilde \rho) = \sqrt{\left(\frac{3}{\Kc}\right)^2+\tilde \rho^2}+\frac{3}{\Kc} \quad \Leftrightarrow\quad h(\rho) = \sqrt{\left(\frac{3}{\Kc}\right)^2+\arcsinh^2\left(\frac{\sinh\rho}{\aconf}\right)}+\frac{3}{\Kc}
h(\tilde \rho) = \sqrt{\aaasquared+\tilde \rho^2}\minusaaa \quad \Leftrightarrow\quad h(\rho) = \sqrt{\aaasquared+\left[\sinh^{-1}\left(\frac{\sinh\rho}{\aconf}\right)\right]^2}\minusaaa
%h(\tilde \rho) = \sqrt{\aaasquared+\tilde \rho^2}\minusaaa \quad \Leftrightarrow\quad h(\rho) = \sqrt{\aaasquared+\arcsinh^2\left(\frac{\sinh\rho}{\aconf}\right)}\minusaaa
\end{equation}
where \eref{compacthyperb} has been used. 
For the compactification factor, we exploit the same functional form \eref{confflat} as for the flat case, namely
\begin{equation} \label{confhyperb}
\aconf(\rho)= \frac{\rho_\scri^2-\rho^2}{2\ \aaa\ \rho_\scri}, %-\Kc\frac{\rho_\scri^2-\rho^2}{6\ \rho_\scri},
\end{equation}
and, same as done before, we set $\rho_\scri=1$.
The final expression for the wave equation is given by the lengthy \eref{ehyperb}, included in the appendix for readability reasons. 

\section{Recovery of non-hyperboloidal time coordinate} \label{section:alg_explained}

\begin{figure}[h]%

\begin{tikzpicture}[scale=7]
        
    % BOUNDING DIAGRAM
    %\node[right, draw] at (0.4,1) {\small{$\tau=t-\sqrt{1+r^2}$}};
    \coordinate (O) at ( 0, 0.06); % center: origin (r,t) = (0,0)
    \coordinate (N) at ( 0, 1); % north: t=+infty, i+
    \coordinate (E) at ( 1, 0); % east:  r=+infty, i0
    \draw[thick] (O) -- (N);
    \draw[thick,dashed] (N) -- (E);

    % LABELS
    \node[right] at (E) {$i^0$};
    \node[above] at (N) {$i^+$};
    %\node[above, rotate=90] at (0,0.2) {\small{$r=0$}};
    \node[above right] at (0.38,0.62) {$\scri^+$};
    \node[left,red] at (0,0.13) {\small $\{ (\tilde T, \tilde r): \tilde T -h(\tilde r) = T_0 \}${\color{black}$\leftrightarrow$}};
    \node[left,red] at (0,0.25) {\small $\{ (\tilde T, \tilde r): \tilde T -h(\tilde r) = T_1 \}${\color{black}$\leftrightarrow$}};
    \node[left,red] at (0,0.37) {\small $\{ (\tilde T, \tilde r): \tilde T -h(\tilde r) = T_2 \}${\color{black}$\leftrightarrow$}};
    \node[left,red] at (0,0.46) {\tiny $\vdots \qquad\qquad\qquad\quad$};
    \node[left,red] at (0,0.51) {\small $\{ (\tilde T, \tilde r): \tilde T -h(\tilde r) = T_n \}${\color{black}$\leftrightarrow$}};
    \node[below,rotate=-30,teal] at (0.65,0.15) {\small$\tilde T = \tilde T_0$};
    
    % COMPACTIFICATION
    \tikzset{declare function={
        T(\t,\r)  = {\fpeval{1/pi*(atan(\t+sqrt(1/4+\r*\r)-1/2+\r) + atan(\t+sqrt(1/4+\r*\r)-1/2-\r))}};
        R(\t,\r)  = {\fpeval{1/pi*(atan(\t+sqrt(1/4+\r*\r)-1/2+\r) - atan(\t+sqrt(1/4+\r*\r)-1/2-\r))}};
        Tspacial(\t,\r)  = \fpeval{1/pi*(atan(\t+\r) + atan(\t-\r))};
        Rspacial(\t,\r)  = \fpeval{1/pi*(atan(\t+\r) - atan(\t-\r))};
    }}
    
    % FOLIATION
    %\message{Drawing time surfaces.^^J}
    \def\Nlines{4} % total number of lines is 2\Nlines+1
    \newcommand\samp[1]{ tan(90*#1) } % for equidistant sampling 
    % hyperboloidal slices
    \foreach \i [evaluate={\t=\i/(\Nlines+4);}] in {1,...,\Nlines}{
        %\message{Drawing i=\i...^^J}
        \draw[line width=0.3,samples=30,smooth,variable=\r,domain=0.001:1,red] plot({ R(\samp{\t},\samp{\r}) }, { T(\samp{\t},\samp{\r}) });
    }
    % spacial slices
    \draw[line width=0.3,samples=30,smooth,variable=\r,domain=0.001:1,teal] plot({ Rspacial(1.,\samp{\r}) }, { Tspacial(1.,\samp{\r}) });
    \draw[line width=0.3,samples=30,smooth,variable=\r,domain=0.001:1] plot({ Rspacial(0.1,\samp{\r}) }, { Tspacial(0.1,\samp{\r}) });
    % dots 
    \foreach \r in {0.,0.37,0.49,0.56} \fill[teal] ({Rspacial(1.,\samp{\r})},{Tspacial(1.,\samp{\r})}) circle (0.01);
    % labels
    \foreach \r in {0.56} \node[above,teal] at ({Rspacial(1.,\samp{\r})},{Tspacial(1.,\samp{\r})}) {\small$1$};
    \foreach \r in {0.49} \node[above,teal] at ({Rspacial(1.,\samp{\r})},{Tspacial(1.,\samp{\r})}) {\small$2$};
    \foreach \r in {0.37} \node[above,teal] at ({Rspacial(1.,\samp{\r})},{Tspacial(1.,\samp{\r})}) {\small$3$};
    \foreach \r in {0.08} \node[above,teal] at ({Rspacial(1.,\samp{\r})},{Tspacial(1.,\samp{\r})}) {\small$n$};
    \foreach \r in {0.26} \node[above,rotate=-20,teal] at ({Rspacial(1.,\samp{\r})},{Tspacial(1.,\samp{\r})}) {\small$\ldots$};
    % dots 
    \foreach \r in {0.56,0.64,0.74} \fill[blue] ({R(0.2,\samp{\r})},{T(0.2,\samp{\r})}) circle (0.01);
    % labels 
    \foreach \r in {0.56} \node[below,blue] at ({R(0.2,\samp{\r})},{T(0.2,\samp{\r})}) {\small$\updownarrow$};
    \node[below,blue] at (0.42,0.23) {\small$ (\tilde T_0,\tilde r_0 )$};
    
    % EXPLANATIONS
    \node[right, draw] at (0.3,1) {\small \begin{varwidth}{16em}Store max value of $\phi$ over intersection points {\color{teal}1,2,3, ... n}\end{varwidth}};
    \draw[->](0.8,0.9) to (0.8,0.75); 
    \node[right, draw] at (0.54,0.65) {\small \begin{varwidth}{9em}Repeat starting from {\color{blue}$\tilde r_1, \tilde r_2, ...$}\end{varwidth}};
    \draw[->](0.58,0.55) to (0.58,0.44); 
    
\end{tikzpicture}

\caption{Schematic representation of how $\sup_{\mathbb{R}^3}|\phi|(\tilde T, \cdot)$ is recovered from data from the simulations on hyperboloidal slices. Given an initial radius $\tilde{r}_0$, there exists a unique $\tilde{T}$--constant level set that intersects the first hyperboloidal slice at $\tilde{r}_0$. Let it correspond to the unrescaled time $\tilde{T}_0$. Then, it is possible to recover the intersections between $\{\tilde T = \tilde{T}_0\}$ and the remaining hyperboloidal slices. If we run the numerical experiment with high enough accuracy, this gives a reliable approximation of the values of $\phi$ in $\{\tilde{T}_0\}\times \mathbb{R}^3$. Therefore, we can store the $L^{\infty}$ norm of $\phi$ at $\tilde{T}=\tilde{T}_0$. Then, we repeat the procedure starting from the radius $\tilde{r}_1 > \tilde{r}_0$ along the first hyperboloidal slice, and so on.}
\label{recovery_diagram}

\end{figure}
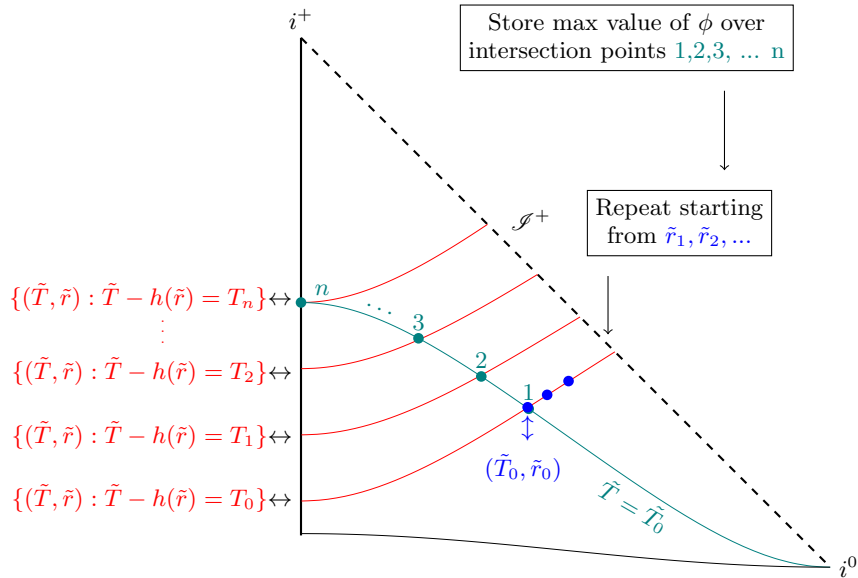

The hyperboloidal approach has its own advantages that we described in the previous sections.  However, for all practical purposes it is desirable to draw a comparison between the numerical results in terms of hyperboloidal slices and those in terms of other spacetime foliations. In the post-processing of data, this is possible whenever the hyperboloidal slices can be covered to a suitable extent by the leaves of the second foliation under comparison. The algorithm we exploited to convert between the hyperboloidal foliation and the $\tilde{T}$-constant slices (where $\tilde T$ is either physical or conformal time, see \eqref{genericttrafo}) is presented in appendix \ref{appendixAlgorithm}, and is schematically described by \fref{recovery_diagram}. 

%%%%%%%%%%%%%%%%%%%%%%%%%%%%%%%%%%%%%%%%%%%%%%%%%%%%%%%%%%
\section{Results: Linear wave equation} \label{linearresults}

We solve for the linear wave equation \eref{wave_eqn} by writing it as a first-order-in-time and second-order-in-space system (see also \cite[section 3]{Rossetti:2023igb}) and by casting the system in terms of the metric coefficients  derived in section \ref{section:hypapproach}. The explicit expressions for the systems we focus on are presented in appendix \ref{eqsappendix}.

Our spherically symmetric code takes advantage of the Method of Lines (4th order finite differences methods were used for radial derivatives on a staggered grid) and the system was integrated in time via a 4th order Runge-Kutta scheme.

As initial data, we take a Gaussian-like pulse centered away from the origin (see \eref{initial_data_system}) and rapidly decaying to zero for large values of the radial coordinate, multiplied by an amplitude factor $\amplitude > 0$. For our numerical purposes, and due to finite machine precision, we regard these initial data as compactly supported.

\begin{figure}[h]%
\centering%
\begin{minipage}{.49\linewidth}%
\centering%
\includegraphics[width=\linewidth]{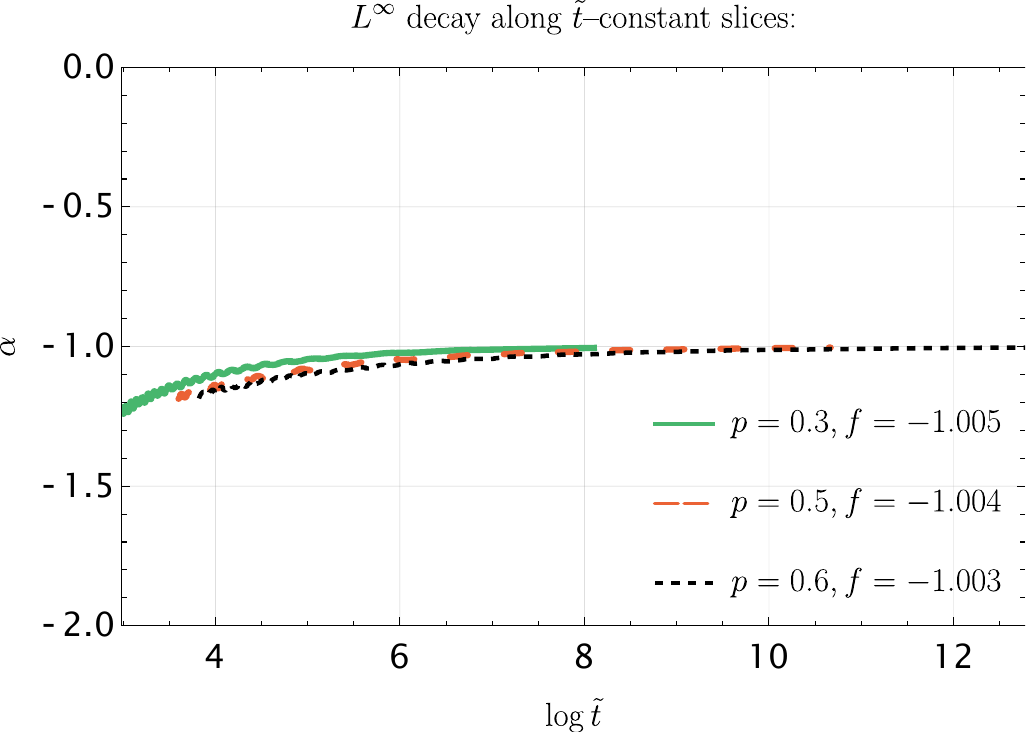}%
\end{minipage}%
\begin{minipage}{.49\linewidth}%
\centering%
\includegraphics[width=\linewidth]{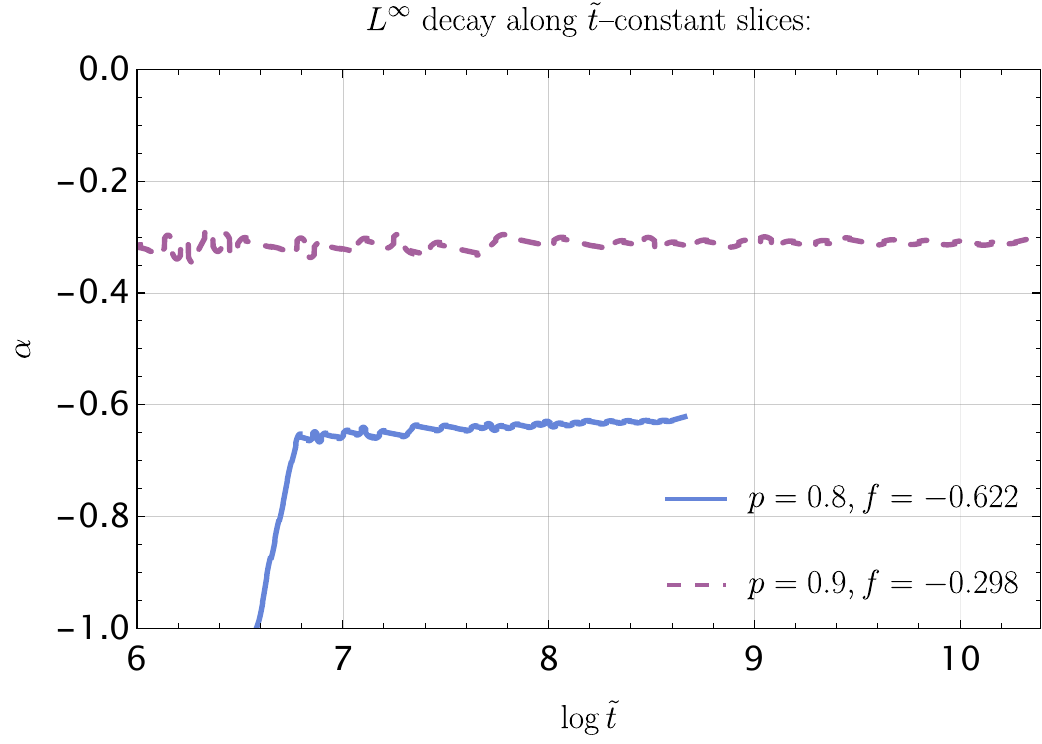}  %\label{fig:decayphiflatplarge}%
\end{minipage}%
\caption{Decay rates for $\phi$ in \textbf{flat} FLRW spacetimes with scale factor $a(t)=t^p$. Here $dt = 0.00025$, $dr=0.00125$, $r \in [0, 1]$. We obtain the expected decay rates, see also \cite{Rossetti:2023igb}. The quantity $f$ measures the final value of the plotted line.}  \label{fig:decayphiflatpsmall}
\end{figure}

\begin{figure}[h]
\begin{center}
\begin{minipage}{.49\linewidth}
\centering
\includegraphics[width=\linewidth]{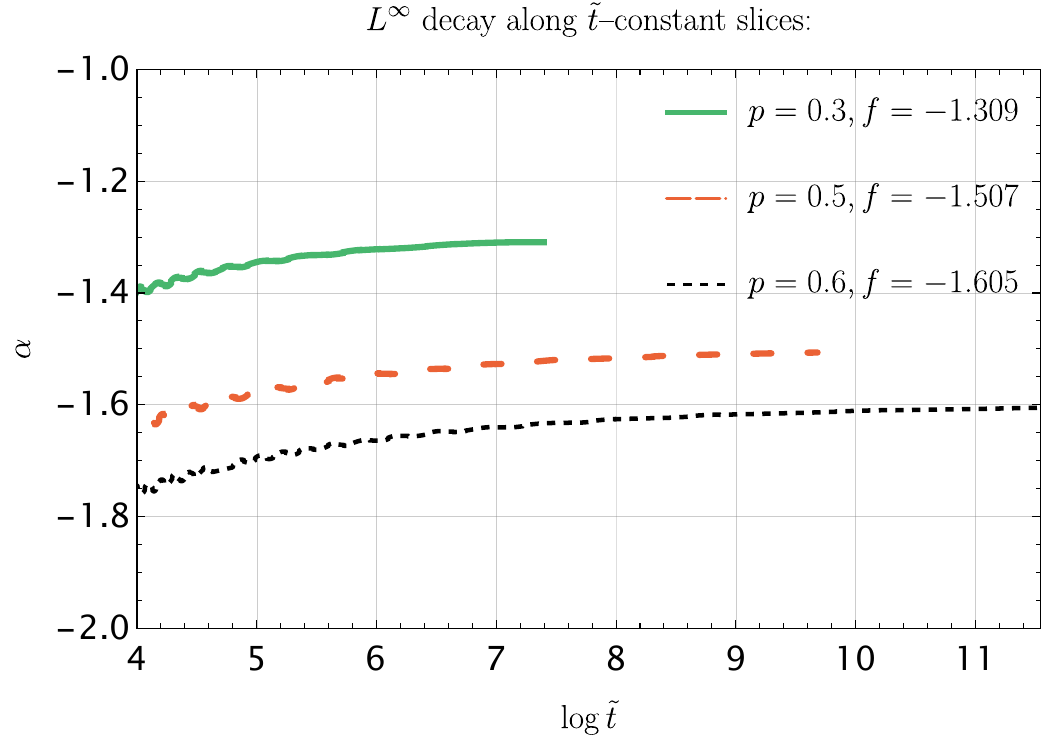}
\end{minipage}%
\begin{minipage}{.49\linewidth}
\centering
\includegraphics[width=\linewidth]{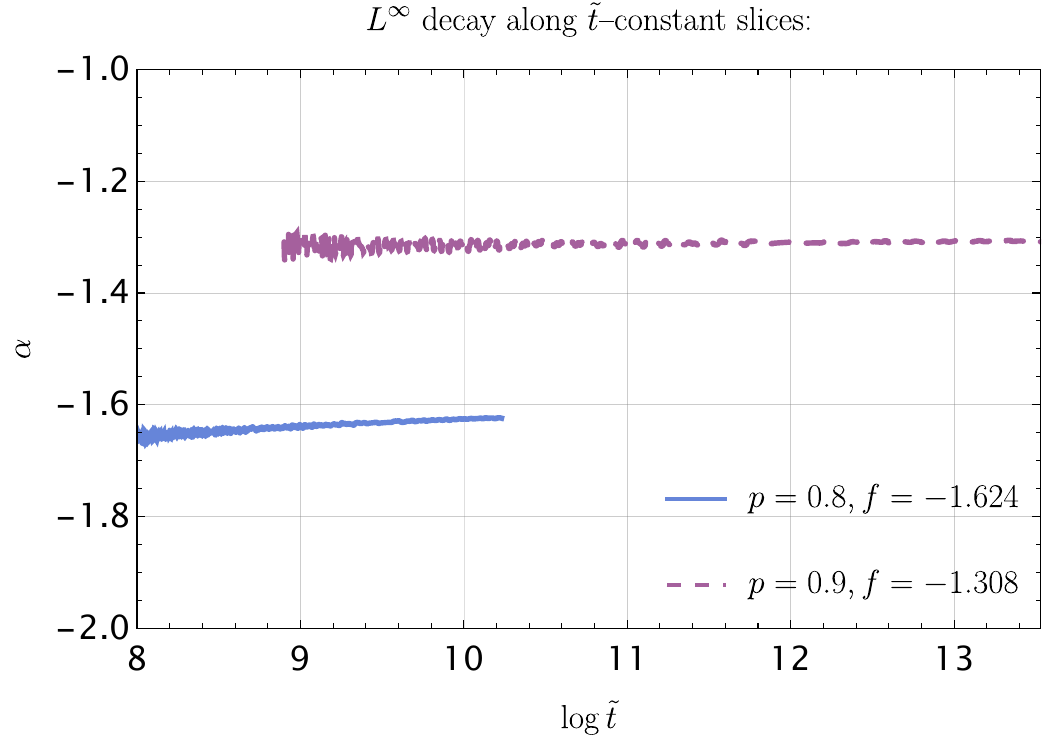}
\end{minipage}
\caption{Decay rates for $\partial_{\tilde t} \phi$ in \textbf{flat} FLRW spacetimes with scale factor $a(t)=t^p$. Here $dt = 0.00025$, $dr=0.00125$, $r \in [0, 1]$. The quantity $f$ measures the final value of the plotted line.} \label{fig:decaydtphiflatpsmall}
\end{center}
\end{figure}

\begin{figure}[h]
\begin{center}
\includegraphics[width=0.7\linewidth]{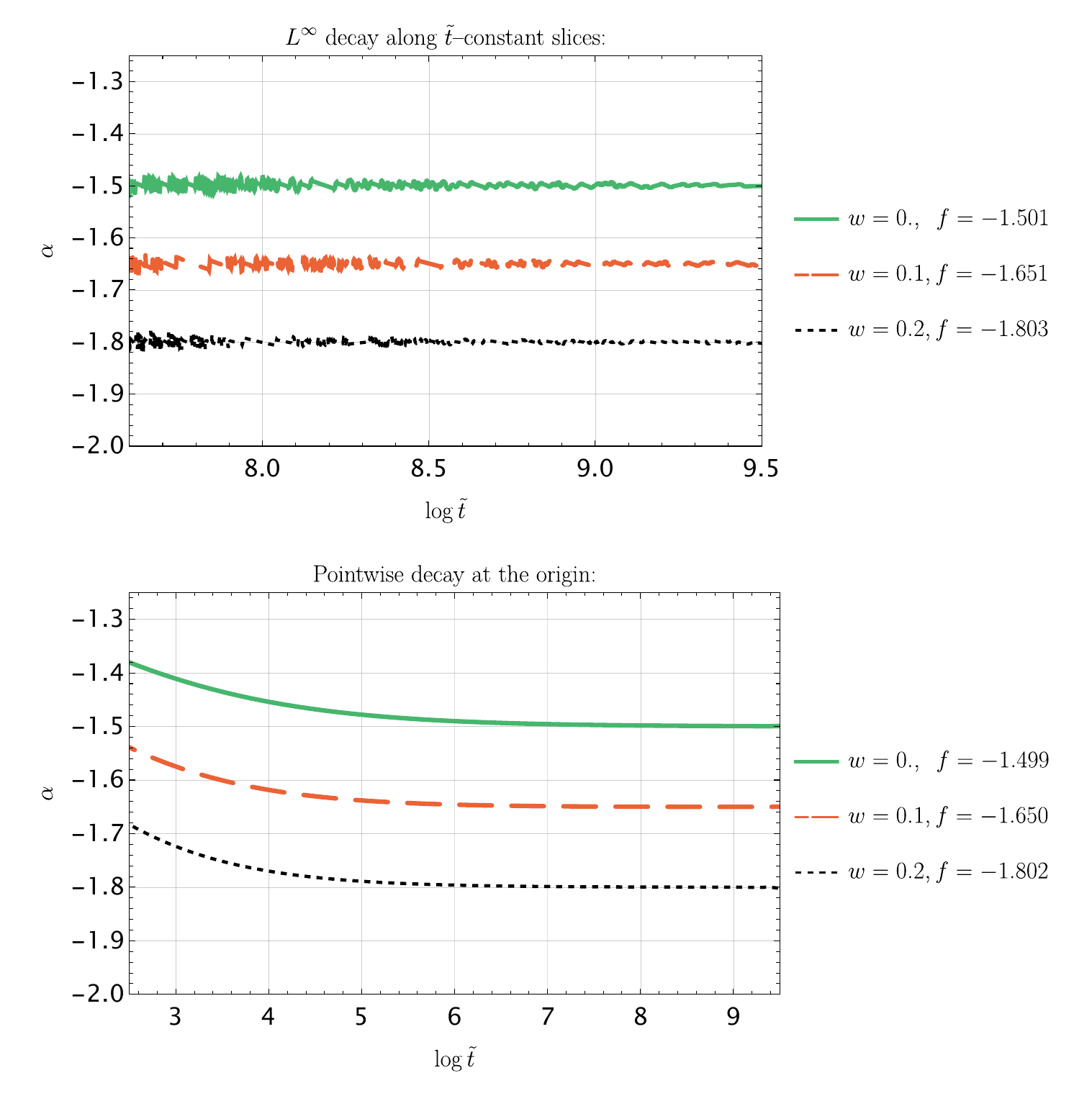}
\caption{Decay rates for $\phi$ in \textbf{hyperbolic} FLRW spacetimes with scale factor as in \eref{scalefactor_hyp} when $0 < w < \frac13$. The $L^{\infty}$ decay rate we obtain along the $\tilde{t}$--foliation comes from the slow decay at the origin. Notice that, at the origin, the value of $\tilde t$ coincides with that of the hyperboloidal time $t$. Here $dt = 0.00025$, $dr=0.00125$, $r \in [0, 1]$. The quantity $f$ measures the final value of the plotted line.} \label{fig:decayphihyperbolicwsmall}
\end{center}
\end{figure}

\begin{figure}[h]
\begin{center}
\includegraphics[width=0.7\linewidth]{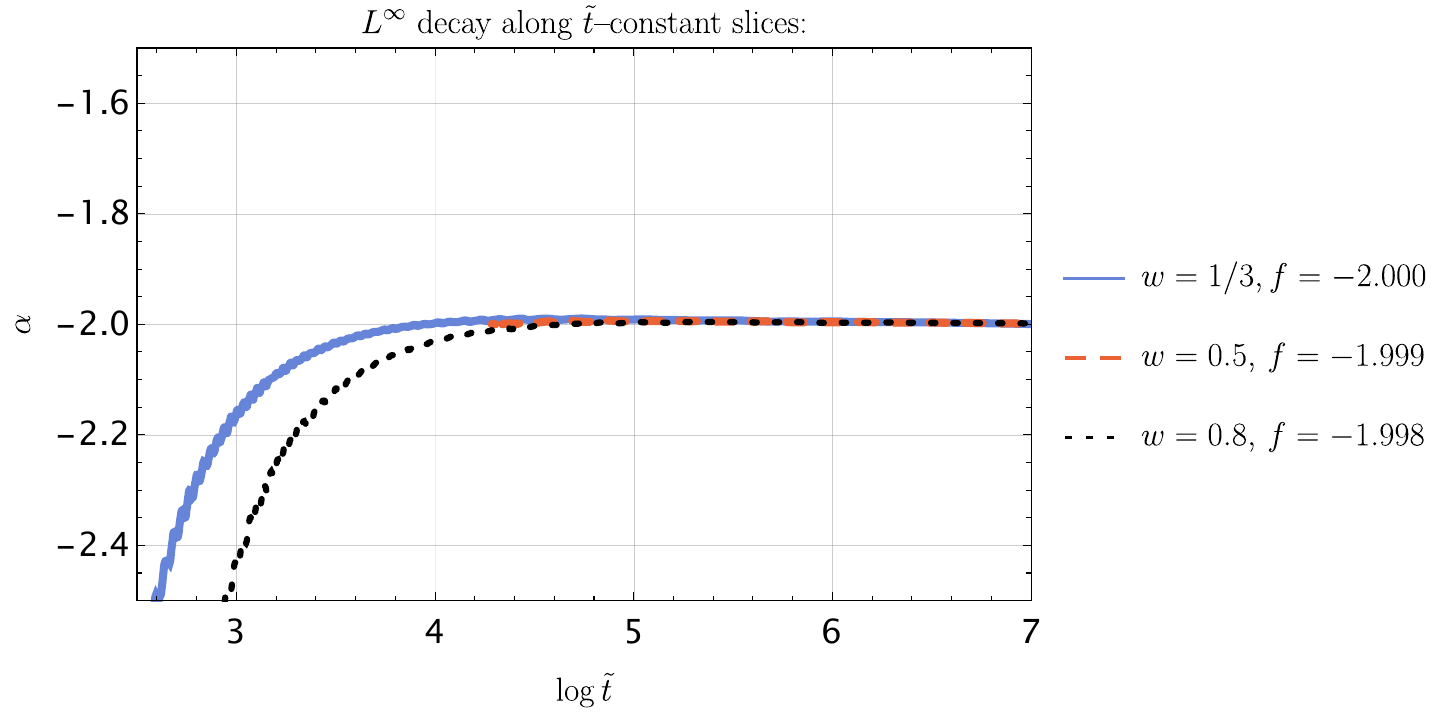}
\caption{Decay rates for $\phi$ in \textbf{hyperbolic} FLRW spacetimes with scale factor $a(t)$ given by  \eref{scalefactor_hyp} when $\frac13 \le w < 1$. The $L^{\infty}$ decay rate we obtain along the $\tilde{t}$--foliation is slower than the decay at the origin.  Here $dt = 0.00025$, $dr=0.00125$, $r \in [0, 1]$. The quantity $f$ measures the final value of the plotted line.} \label{fig:decayphihyperbolicwlarge}
\end{center}
\end{figure}

The decay rates we represent in the  following plots are given in terms of a parameter $\alpha < 0$.\footnote{Not to be confused with the lapse $\alpha$ appearing in appendix \ref{eqsappendix}.} In particular, since $\phi$ is expected to decay as $\phi \sim \tilde{t}^{\alpha}$ (\textbf{all decay rates are given in terms of physical, unrescaled, time} to compare with known results), we plot $\frac{d \log \phi}{d\log \tilde t} \sim \alpha$ as a function of $\log \tilde t$. See also Figs. \ref{fig:decayphiflatpsmall}, \ref{fig:decaydtphiflatpsmall}, \ref{fig:decayphihyperbolicwsmall}, \ref{fig:decayphihyperbolicwlarge}, where we employed the algorithm described in section \ref{section:alg_explained} to convert the decay results between the $t$ and the $\tilde t$ coordinates. 

In the spatially--flat case given by metric \eref{FLRWflatmetric}, with 
\[
a(\tilde t)= \tilde t^p,
\]
we retrieve the same decay results for $\phi$ and $\partial_{\tilde t} \phi$ that we previously obtained in \cite[eqns. (11) and (12)]{Rossetti:2023igb} for $0 < p < 1$, see also Fig. \ref{fig:decayphiflatpsmall} and Fig. \ref{fig:decaydtphiflatpsmall}. Compared to the non--hyperboloidal approach of \cite{Rossetti:2023igb}, the final decay results are represented in terms of non--smooth plotted lines due to the conversion algorithm described in section \ref{section:alg_explained}. As expected we obtain smoother lines when the numerical output is stored in memory more frequently and/or we set smaller values of $dr$.

%We did not carry the same analysis for larger values of $p$, but we are confident that similar results hold provided that the hyperboloidal foliation is modified so to take into account the spacelike character of \scrip. 

Similarly, we found consistent decay rates in the spatially--hyperbolic case, with scale factor
\begin{equation} \label{scalefactor_hyp}
a(\tilde \tau) =\left [ \sinh \left ( \frac{3w+1}{2} \tilde \tau \right ) \right ]^{\frac{2}{3w+1}},
\end{equation}
where $0 \le w < 1$ and $\tilde \tau$ is the conformal time (we recall that, as a consequence of the Friedmann equations, we have decelerated expansion for $w > -\frac13$.). We also refer to Fig. \ref{fig:decayphihyperbolicwsmall} and Fig.  \ref{fig:decayphihyperbolicwlarge}, from which it is clear that the $L^{\infty}$ decay for $0 \le w \le \frac13$ is driven by the slow contribution at the origin of the numerical grid. On the other hand, when $\frac13 < w < 1$, the pointwise decay at $\tilde r = 0$ is faster than the decay in the supremum norm.

Some comments about the framework we exploit are in order:
\begin{itemize}
\item The construction given by the hyperboloidal foliation is such that we do not have to deal with spurious contributions coming from the boundary conditions at infinity. This  allows us to evolve for effectively larger evolution times.
\item For each value of $p$, we found a (large) maximum time until which numerical experiments are reliable: near such a time value, convergence fails due to the blow-up of linear solutions. This time limitation depends on the value of $\kappa$ (see \eref{kappaK}), i.e. depends on the curvature of the hyperboloidal slices (see e.g. \cite[Fig. 3.1]{VVthesis} for Penrose diagrams corresponding to different values of $\kappa$). In particular, 
for $0 < p < 1$, $p$ close to 1 (i.e.  for a fast expansion of the underlying universe) we found the simulations to crash very quickly when $\kappa = 1$. Larger values such as $\kappa = 3$ allowed to evolve for longer. In short, we found that, in order to run the experiments for a sufficiently long time, an inverse proportional relation between the rate of the expansion of the universe and the curvature of the hyperboloidal slices has to be respected.
\item We found the above decay rates by running the experiments for an unrescaled solution $\phi$. Similar results were obtained  when rescaling solutions by a factor ${\Omega}^{\gamma}$, for specific choices of $\gamma \in \mathbb{R}$. %{\fr \sout{, provided that no degenerate behaviour is introduced in the main  coefficients in the partial differential equations under scrutiny}}. {\avv @Flavio: What do you mean with degenerate behaviour?}{\fr Flavio: I meant growing behaviour at infinity, however see the corrected sentence.}
\item We  ran the simulations by exploiting both the height function of section \ref{subsec:flatconformal} (which is used for the plots we display in this work) and its time--dependent version introduced in section \ref{subsection:flatphysical}. The latter gives comparable results with respect to the former, however it requires much longer evolution times for large values of $p$. Indeed, the time--dependent height function gives rise to level sets that reach higher values of $\tilde u$ more slowly when $p > \frac12$. However, we believe that this phenomenon can be controlled by making a suitable choice of $\C(t)$ (see section \ref{subsection:flatphysical}), on which little constraints are imposed. Considering a time--dependent value of $\kappa$ is also admissible, but this requires a time--dependent compactification factor and goes beyond our simpler construction.
\end{itemize}

%{\avv general comments on experiments/results by Alex:
%\begin{itemize}
%\item Flat: as Flavio said, decreasing the absolute value of $\Kc$ is needed to run the simulations with large $p$ for long enough before they crash at scri. This needs to be mentioned (time limitation). Does this also happen for smaller $p$, but we have not detected it? Update: above a certain value of $\Kc$ the simulations seem to run forever (checked with $p=0.5$, $\Kc=-3.5$ ($\Kc=-3.8$ crashed) stable until t=50). So all good, just mention the minimum alue of $\Kc$ required for stability, which increases with higher values of $p$. 
%\item Shall we try that with most aggressive rescaling $\aconf^{\frac{1}{1-p}}$ too? Or is it ``unconditionally unstable''?
%\item Have decay rates been recovered using Flavio's height function? Do we need to run for longer, as the evolution seems to be slower? 
%\end{itemize}
%}

%\subsection{Comparison to non-hyperboloidal results}

% By employing the hyperboloidal approach, we obtained the following decay rates for solutions to the wave equation in flat FLRW spacetimes, in complete accordance with the results of [math\_paper] and the numerical results [num\_paper] that suggested that such decay rates are sharp%

%%%%%%%%%%%%%%%%%%%%%%%%%%%%%%%%%%%%%%%%%%%%%%%%%%%%%%%%%%
\section{Results: Non-linear wave equations} \label{section:nonlinearwave}
%{\fr Flavio: in the last version I swapped $-c_2$ with $c_2$. Say that it is intrinsically difficult due to the loss of convergence near the blow-up. A possible question is: how do we know if the blow-up is caused by $K_{CMC}$? Notice that, for $p$ small, $Kcmc$ does not matter at the linear level. Moreover, we get comparable results when using a different foliation.}
Given semi--linear wave equations of the type 
\begin{equation} \label{wave_nonlin}
\square_{g} \phi = c_1 a(\tilde t)^{\beta_1} \left(\partial_{\tilde t} \phi \right)^2 + c_2 a(\tilde t)^{\beta_2} \left(\partial_{\tilde r} \phi \right)^2 + c_3 a(\tilde t)^{\beta_3} \left (\partial_{\tilde t}{\phi} \partial_{\tilde r}{\phi} \right)
\end{equation}
with  $c_i \in \mathbb{R}$, $\beta_i \in \mathbb{R}$, $i = 1, 2, 3$, we want to verify the existence of global--in--time solutions when our initial data are sufficiently small in $L^{\infty}$ norm:
\begin{subequations}\label{initial_data_system}
\begin{eqnarray}
\phi(\tau_0, r) &=&\amplitude \ G ( r) \label{nonlin_initial_data} \\
%{\fr\sout{ \partial_{\tilde \tau }\phi(\tilde{\tau}_0, r)}} = 
\partial_{\tau}\phi(\tau_0, r) &=&  -\frac{\amplitude r }{\kappa} \ \partial_r G( r)
\end{eqnarray}
\end{subequations}
for $\amplitude$ small and with
\[
G(x) = \exp \left(-\frac{(x^2-c^2)^2}{4 \sigma^4} \right), \quad c \ge 0, \sigma > 0.
\]
The equations implemented in the numerical code are those described in appendix \ref{eqsappendix_conformal} for the linear case, where appropriate non--linearities are added according to the scenarios analysed in the following paragraphs.
 For instance, partial derivatives $\partial_{\tilde t}$ are turned into $\partial_\tau$ and $\partial_r$ using the chain rule with $d\tilde t = a\ d\tilde\tau$, \eref{tautrafo} and \eref{compactflat}.

Small data global existence for \eqref{wave_nonlin} cannot hold for all parameters $\beta_i$ and $c_i$, $i=1, 2, 3$, since, in Minkowski spacetime, already the $c_1=1,\,c_2=c_3=0$, $\beta_1 = \beta_2 = \beta_3 = 0$ case, i.e.
\begin{equation} \label{fritzjohn_example}
\square_{g} \phi = \left(\partial_{\tilde t} \phi \right)^2,
\end{equation}
 presents a blow-up behaviour in finite time for every non--trivial solution \cite{Sogge}. On the other hand, a null condition exists in Minkowski and, in flat FLRW for $p>1$, a null condition was found in \cite{CostaFranzenOliver}  to be given by the right hand side of \eref{wave_nonlin} with $\beta_1 = 0,  \beta_2 = -2, \beta_3 = -1$ and any value of $c_1, c_2, c_3$.\footnote{The global existence result in fact holds if the constants $c_i$, $i=1, 2, 3$ are replaced by more general smooth and bounded functions $\mathcal{N} = \mathcal{N}(\phi)$ having uniformly bounded derivatives of all orders.} These choices for $\beta_i$, ${i=1, 2, 3}$ are motivated by the decay of the time derivatives (so that the non-linear terms decay faster than $\tilde{t}^{-2}$).\footnote{Whereas in  Minkowski spacetime the null condition, rather than depending on decay only, is a stricter constraint on the bilinear forms describing the non--linear terms. When the null condition holds, the quadratic non--linear terms in the wave equation decay as $\tilde{t}^{-3}$.}
 
At the numerical level, investigating small data global existence presents several difficulties:
\begin{enumerate}
\item Even for the ``good'' non--linearities, i.e.\ those yielding small data global existence, it is always possible to obtain a blow--up behaviour if the initial data are large. Since there is no a priori scale for the initial data, it is essential to distinguish between the blow-up related to large data and the blow-up related to ``bad'' non--linear terms.
\item In the case of the ``bad'' non--linearities, i.e.\ those not giving small data global existence, an almost--global existence result is expected in light of the established theory in Minkowski spacetime \cite[theorem 2.2]{Sogge}. The notion of almost--global here denotes the dependence of the time of existence of solutions, $T=f(A)$, on a function $f > 0$ such that $f(A) \to +\infty$ as the factor $A$ multiplying the initial data approaches zero. It is therefore essential to distinguish between an infinite time of existence and a time of existence which is finite but ``very large'' due to the smallness of the initial data.
\item Near the blow--up time, loss of convergence is expected due to the diverging behaviour of solutions. It is essential to distinguish between the failure of convergence due to the blow-up and that due to external factors, e.g. the value of $\kappa$ (see section \ref{section:hypapproach}).
\end{enumerate}
% $\tilde t^* = \tilde t^*(\epsilon)\sim e^{\frac{c}{\epsilon}}\sim\frac{c}{\epsilon^2}$, where $\epsilon$ controls the ``smallness'' of the initial data
All the above considered, we proceed as follows. First, we study the behaviour of solutions for different non-linear terms in the case of Minkowski spacetime (i.e. a flat FLRW spacetime with $p=0$), for which results are known. We verify that, despite the difficulties we listed,  the case of a ``bad'' non--linearity can be clearly distinguished from that of small data global existence, solely by analysing the numerical output we obtain. 
In particular, let $\amplitude_{\text{Max}}$ be the largest size of the initial data\footnote{To be precise, $\amplitude_{\text{Max}}$ is the amplitude that multiplies the Gaussian-like initial data, see \eref{nonlin_initial_data}.} for which the solution exists (and three runs of convergence tests  with 3x resolution increase are successfully completed) until a large hyperboloidal time $T$. Such an $\amplitude_{\text{Max}}$ can be found for every choice of non--linear terms, due to the above remarks. Then, we analyse the consequences of a slight increase in the size of the initial configuration. We are interested in the smallest $\delta_{B.U.}$  (with ``B.U.'' denoting blow-up) such that initial data of size
\[
\amplitude_{\text{Max}}(1 + \delta_{\text{B.U.}})
\]
lead to divergence in $L^{\infty}$ norm at earlier times. We identify the blow-up behaviour with an overflow in the code, and we denote the time of existence associated to the value of $\delta_{\text{B.U.}}$ by $T^*$, where $T^* \le T$. When the null condition holds in Minkowski spacetime, the value of $\delta_{\text{B.U.}}$ is considerably smaller than that in the case of the Fritz John example \eref{fritzjohn_example}. This is in line with the fact that, in the latter case, an almost--global existence result holds and thus the time of existence grows, roughly speaking,  ``continuously'' with respect to the size of the initial data. The scenario is quantitatively different from  the case of the null condition, where heuristically we expect a discontinuity in the time of existence as a function of the initial configuration.

\begin{figure}[h]
\begin{center}
\includegraphics[width=0.7\linewidth]{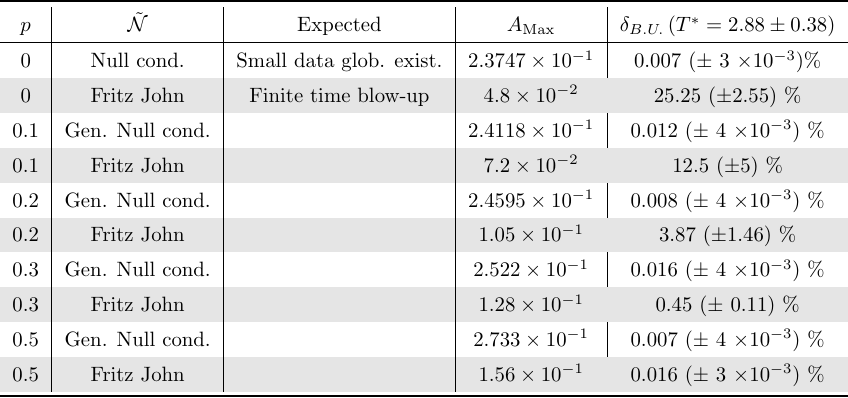}
\caption{$\amplitude_{\text{Max}}$ is the ``maximum'' size of data for which we have existence until hyperboloidal time $10$: up to such value, the solution exists, is bounded and three runs of convergence tests can be completed. For initial data of size $\amplitude_{\text{Max}}(1 + \delta_{B.U.})$, overflow occurs for either $\phi$ or its time derivative at time $T^* < 10$. For even larger data, the solution also blows up and the respective $T^*$ becomes smaller. Here we list the values of $\delta_{B.U.}$ for different non-linear terms $\tilde{\mathcal{N}}$, where we considered $dr=0.00125$, $r \in [0, 1]$, $dt = 0.00025$.}  \label{table:resultsNoHyp}
\end{center}
\end{figure}

We repeat this analysis for flat FLRW spacetimes endowed with $0 < p < 1$ (see table \ref{table:resultsNoHyp}) and compare with the results for the case $p=0$.
 
The results show that, when a generalized null condition is taken into account, namely
\begin{equation} \label{gen_null_cond}
\square_{g} \phi = \left(\partial_{\tilde t} \phi \right)^2 -  \frac{1}{a(\tilde t)^{2}} \left(\partial_{\tilde r} \phi \right)^2,
\end{equation}
the value of $\delta_{\text{B.U.}}$ is comparable to that of the case $p=0$ (Minkowski spacetime) when the null condition is implemented. This \textbf{suggests} that \textbf{small data global existence holds} for solutions to \eref{gen_null_cond} for every $p$ with $0 \le p < 1$.

\begin{figure}[h]%
\centering%
\begin{minipage}{.45\linewidth}%
\centering%
\includegraphics[width=0.95\linewidth]{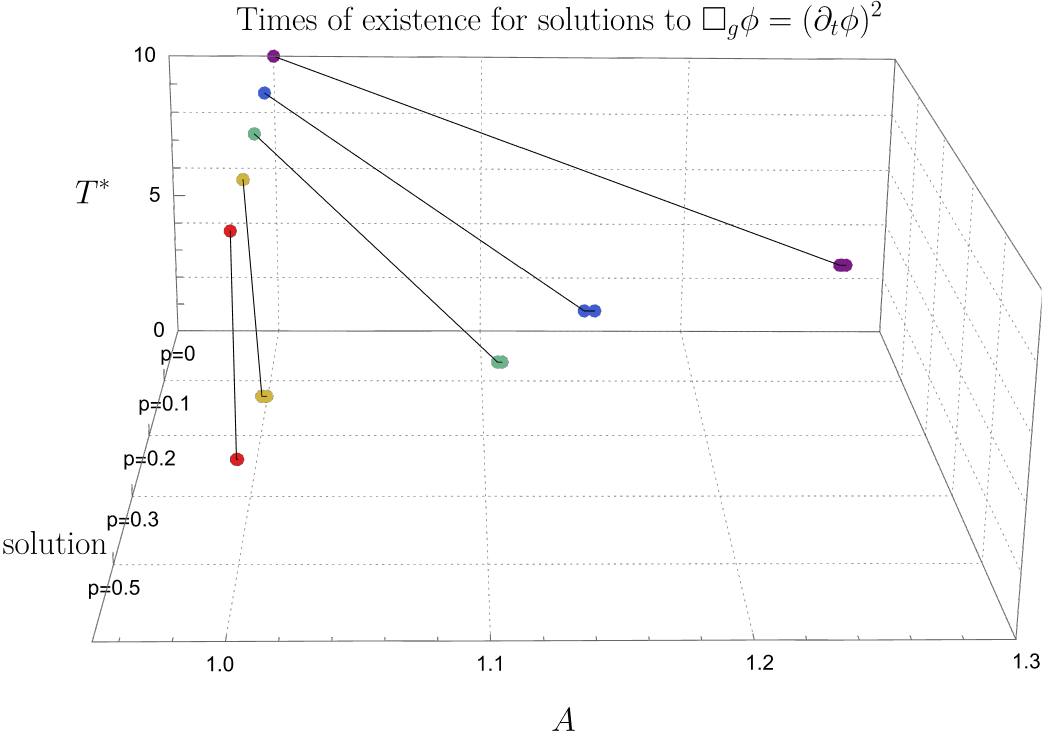}%
\end{minipage}%
\begin{minipage}{.45\linewidth}%
\vspace{1em}
\centering%
\includegraphics[width=\linewidth]{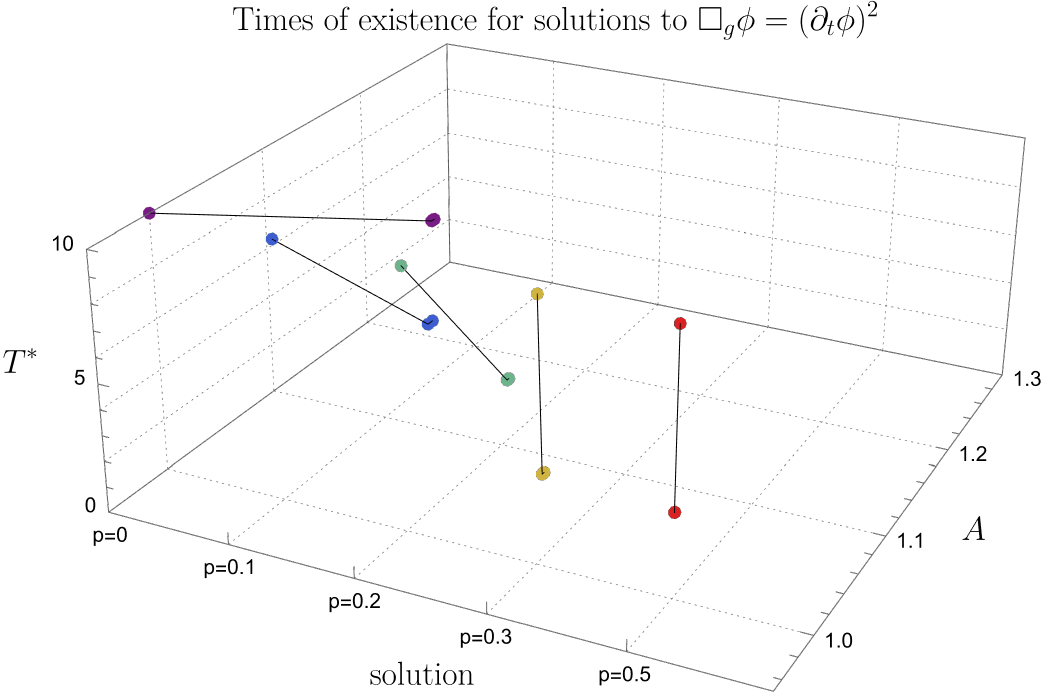}  
\end{minipage}%
\caption{Two different views of the same 3D plot. Along the vertical axis the time of existence of solutions is represented, as a function of the initial data, for Fritz John equations for different values of the rate of expansion $p$. The data were normalized so that, for all solutions, $A_{\text{Max}}=1$ is the maximum value of the initial data yielding the (hyperboloidal) time of existence $10$. At such time value, the initial configuration has already crossed the  outer boundary of the numerical domain. For different values of $p$, the function $T^*$ has different slopes. An infinite slope is identified with a ``discontinuity'' of $T^*$, which is what can be observed in Minkowski spacetime when the null condition holds. This is the behaviour we detect for large values of $p$.}   \label{Fig:3Dplot}
\end{figure}

For the Fritz John example \eref{fritzjohn_example}, we observe a diverging behaviour similar to that of the Fritz John equation in Minkowski, at least for small values of $p > 0$. However, the scenario is different when $p \ge \frac12$. In particular, we observed that the value of $\delta_{\text{B.U.}}$, in the Fritz John case \eref{fritzjohn_example}, decreases as $p$ increases. In other words, this \textbf{suggests} that \textbf{small data global existence fails} for the Fritz John equation for slow spacetime expansions, but it \textbf{might hold} for faster (decelerated) spacetime expansions, namely for $p \ge \frac12$. We also refer to Fig. \ref{Fig:3Dplot} for a visualization of our results and where the discontinuity of the blow-up time as a function of the initial data can be visually associated to the Fritz John case when $p$ is sufficiently large. 

We emphasize that our numerical investigation dealt with the decelerated case $0 < p < 1$, for which rigorous results on small data global existence for semi-linear wave equations are not generally available. The accelerated scenario ($p > 1$) was rigorously studied in \cite{CostaFranzenOliver}.
 
A related problem in FLRW spacetimes is that of studying the dynamics of a test  fluid, rather than a scalar field, described by the Euler equations. This can be seen as a proxy for the Einstein--Euler system.
Such a test fluid  is distinct from the perfect fluid permeating the FLRW spacetimes we consider, since the latter is constrained by the relation $p=\frac{2}{3(1+w)}$ (i.e. its speed of sound is restricted by the expansion rate of the FLRW background) and is described by the Friedmann ordinary differential equation system, to which the Euler equations reduce due to  symmetry. Despite the differences, it is interesting to notice that there is an analogous threshold ($ w= \frac13$, with $w$ being the square of the speed of sound of the fluid) between stability and instability results for the relativistic Euler equations on a FLRW background, see e.g. the recent review \cite{FajmanOfnerWyatt}.

Finally, we stress that, due to numerical constraints, even though solutions to the equations we consider exist for a long time $T$, we cannot exclude their blow-up (caused either by the theory or by numerical limitations) for larger times. Still, our findings are consistent with additional numerical data, obtained via non--hyperboloidal methods, which we will present in forthcoming work.

%\subsection{Experiments}

%use $||\phi||_{L^\infty}$

%1. check $a=1,\,b=0$ and $a=1=b$ for Minkowski (if desired, can use the rescaling $\phi = \frac{\phi}{\Omega}$)

%3.  Something José made me notice: when $p>1$ the Penrose diagram of the FLRW metric looks like that of de Sitter (\scrip \ is spacelike). How does the hyperboloidal foliation look like in that case? See \fref{sketch} for hypothesis. Will be interesting to look into this.
%\begin{figure}[h]
%\begin{center}
%\includegraphics[width=0.4\linewidth]{fig/p_greater_1.png}
%\caption{Sketch for flat FLRW $p>1$ case, whose \scrip is spacelike as for AdS, with hypothetical hyperboloidal slices in red. Worry not, Flavio, this won't go to the journal.}
%\label{sketch}
%\end{center}
%\end{figure}

%4. do hyperbolic FLRW

\section{Convergence}\label{sec:convergence}
\begin{figure}[h]%
\centering%
\begin{minipage}{.45\linewidth}%
\centering%
\includegraphics[width=\linewidth]{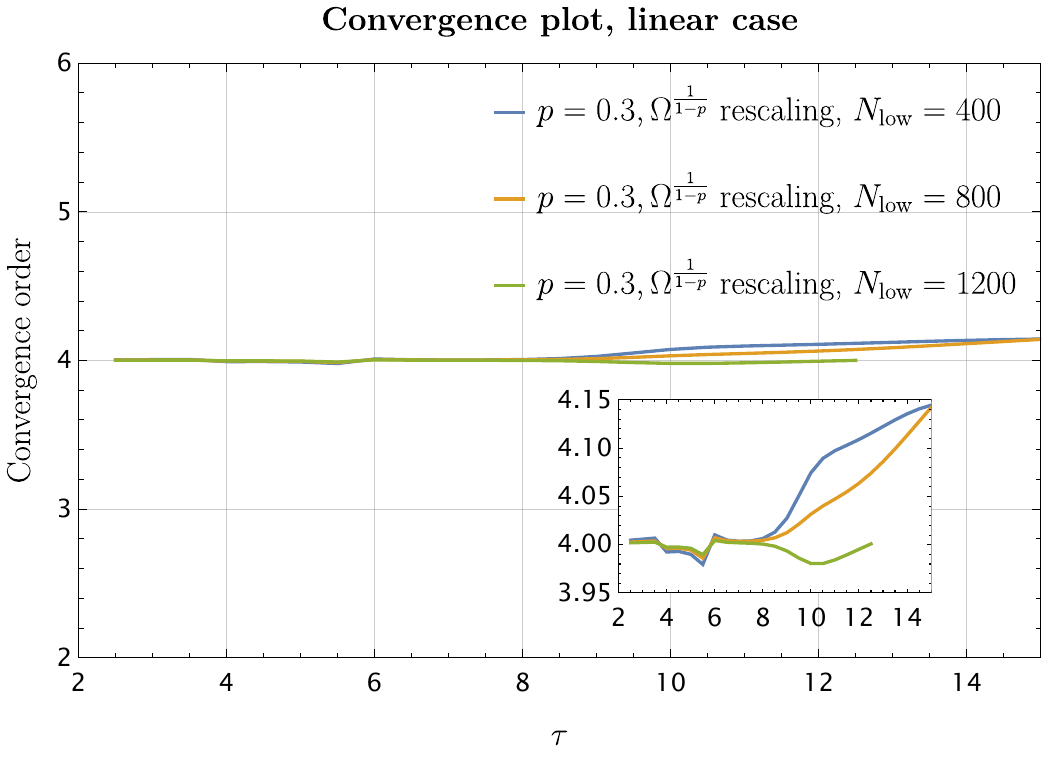}
\end{minipage}%
\hspace{1em} \begin{minipage}{.45\linewidth}%
%\vspace{1em}
\centering%
\includegraphics[width=\linewidth]{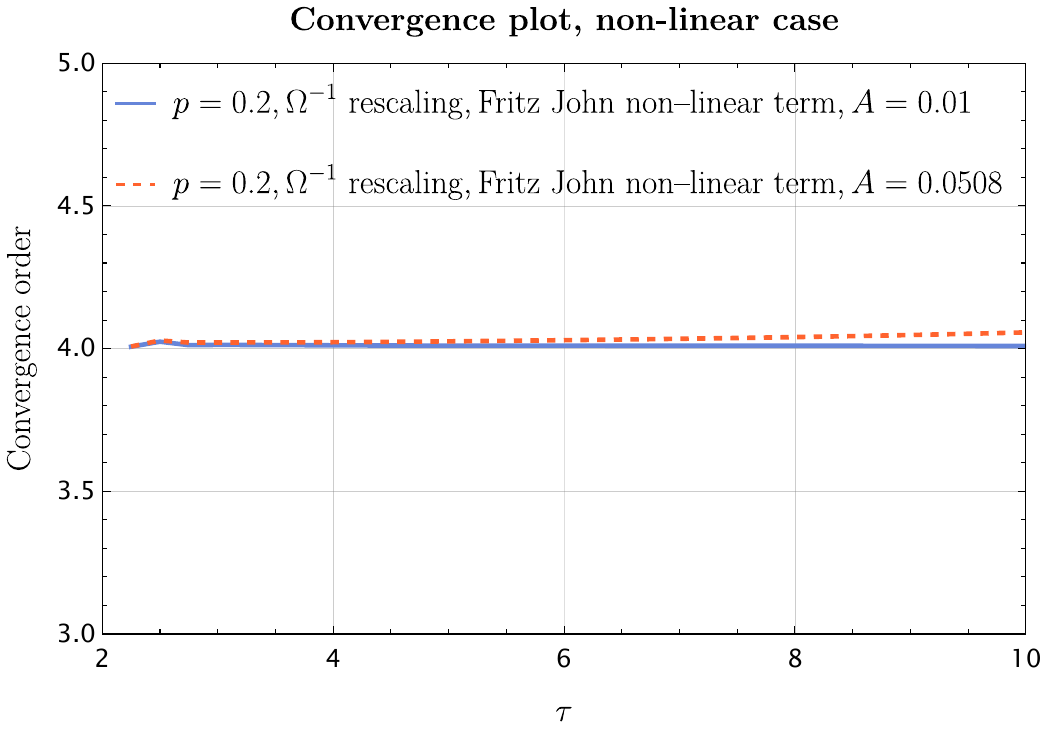}  
\end{minipage}%
\caption{ Convergence order over time using the $L^2$-type norm as $\log_f\left(\frac{\sqrt{\sum_{i=1}^N(\phi_{low, i}-\phi_{med, i})^2}}{\sqrt{\sum_{i=1}^N(\phi_{med, i}-\phi_{high, i})^2}}\right)$ with $i\in[1,N]$ unless otherwise specified. The increase in resolution is $f=3$, the number of points $N_{low}=400,800,1200$  respectively for each of the series, and $\phi_{low/med/high, i}$ denotes the scalar field for the low, medium and high resolutions at point $i$. In the linear case, we observe  very good 4th order convergence. The accuracy increases for higher values of $N$.  The green curve is shorter because this was the most expensive simulation set and we had to stop it before being able to reach $\tau=15$. Similar results are obtained in the non--linear case whenever the initial data are sufficiently small so to have a long time of existence. We tested both the case $A=0.0508$, which is the maximum size of initial data giving global existence and convergence until $\tau = 10$ (for $p=0.2$ and $\Omega^{-1}$ rescaling of the scalar field) and the case of smaller initial data. In the latter case, convergence is even closer to the expected 4th order. } \label{Fig:convergence}
\end{figure}

As a check of the reliability of our code, we obtained the convergence plots in Fig.~\ref{Fig:convergence} for the linear and non-linear wave equations in the case of flat spatial sections and time-independent height function. 

When studying the convergence of our numerical solutions, we observed the following:
\begin{itemize}
\item In the linear case, and in the non-linear case in absence of blow-up, all solutions converge. However, proper 4th-order norm convergence is only found when evolving the scalar field rescaled by a suitable factor (e.g. $\aconf^{\frac{1}{1-p}}$ in the linear case and $\aconf^{-1}$ in the non-linear case), see also Fig. \ref{Fig:convergence}. 
\item  The convergence runs of the linear wave equation were performed with quadruple precision (using 16 bytes to store values instead of the 8 bytes of the usual double precision), as otherwise convergence was not correctly recovered for the series with larger $N_{low}$.
\item For unrescaled scalar fields, 4th-order convergence can be retrieved when restricting the analysis of convergence to a subset of the spatial grid away from $r=1$, i.e. away from \scrip. This is related to the need of rescaling: indeed, unrescaled scalar fields are  exactly zero at \scrip\xspace and thus the convergence errors there are relatively large compared to the size of the analytical solution, which we believe is what hinders convergence.
\item When looking at pointwise convergence results, we observed that the difference between our numerical solution and the analytical one at \scrip\xspace is roughly $10^{-10}$ times smaller than the size of the initial data. 
\item The observed slight loss of convergence due to the presence of \scrip\xspace  for unrescaled scalar fields does not affect our decay results, which depend on the $L^{\infty}$ norm of solutions rather than on the local behaviour at \scrip. The unrescaled scalar field is asymptotically zero at \scrip, so for our purpose of studying the late-time decay we will never measure it exactly there.
\item Results for the non--linear evolutions are also unaffected. We stress, however, that different rescalings for solutions lead to different times of existence in the case of ``bad'' non--linearities.
\end{itemize}

%Proper expected 4th-order norm convergence is found evolving the scalar field rescaled by $\aconf^{\frac{1}{1-p}}$, as can be seen on the left of \fref{Fig:convergence}. This was run with quad precision. Just including the scalar field. 
%For simulations with the unrescaled scalar field such a flat 4th order norm convergence could not be obtained if including the differences all the way to \scrip. However, if only gridpoint up to $r\sim0.93$ are considered, then the norm convergence looks perfect. 
%when solutions become very close to zero in the evolution, convergence looks worse as the errors are proportionally larger. This is why we believe the rescaled scalar field at \scrip gives better convergence than the unrescaled one. 
%we are being very conservative here (mention pointwise convergence plot. )
%The results of this paper are unaffected by these convergence results: we are interested in the decay of the field, and as the unrescaled scalar field is zero at \scrip, then we are never measuring it there and thus any potential imperfections in the convergence will not alter out results. 
%Not providing plots for the setup with time-dependent height function or the spatially hyperbolic case because the numerical setup is exactly the same.
%+ comments for the non-linear case. 

%%%%%%%%%%%%%%%%%%%%%%%%%%%%%%%%%%%%%%%%%%%%%%%%%%%%%%%%%%
\section{Conclusions}

In this work, we analysed solutions to linear and non-linear wave equations propagating in FLRW spacetimes characterized by a decelerated expansion of the non--compact spatial sections. In the linear case, we provided further evidence for the sharpness of decay rates in flat and hyperbolic FLRW solutions. In the non-linear case, our numerical experiments suggest that a natural choice for the null condition should yield small data global existence for semi--linear wave equations in flat FLRW spacetimes. The behaviour of solutions to the Fritz John equation in these spacetimes was also systematically described in the unexplored regime included between the case $p=0$ (Minkowski spacetime) and the case $p>1$ (accelerated expansion). In particular, we found that, in the decelerated FLRW regime, a large expansion rate seems necessary in order to have small data global existence for the Fritz John equation. A further investigation of this phenomenon is left to future numerical and mathematical work.

To the best of our knowledge, this is the first time that hyperboloidal methods are applied to FLRW backgrounds.
Our results were achieved via a modern numerical approach based on a hyperboloidal foliation of the spacetime, which proved to be advantageous due to its feature of encoding information up to \scrip\xspace in slices corresponding to finite  coordinate time values. The hyperboloidal method was specifically adapted to the expansion of the underlying spacetimes. General time--dependent height functions were also constructed. We expect to be able to implement them in a more comprehensive setup by imposing
reasonable constraints on the metric components and solving the resulting differential equation for $h$.

% To the best of our knowledge, this is the first time that the hyperboloidal method is applied to a background spacetime that is not vacuum (while being asymptotically flat). 
Depending on the FLRW model, an infinite amount of e.g.~dust or radiation is compressed on each compactified hyperboloidal slice. It is a fair question whether this compression effect is accountable for any potential numerical challenges, such as maybe the convergence issues at \scrip\xspace described in \sref{sec:convergence}. We plan to further investigate this in the future.  

We hope that our results can offer a new perspective on the modern  mathematical and numerical challenges involved in the study of linear and non--linear problems in FLRW spacetimes. Moreover, the hyperboloidal approach we employed is very general. We are confident that its applications can be extended to a wide range  of numerical studies related to hyperbolic systems in cosmological settings.

%\ack
\section*{Acknowledgements}

The authors thank João Costa, Edgar Gasperin, David Hilditch, José Natário and Jesús Oliver for useful comments on the work and on a preliminary version of the manuscript, and Christian Peterson for physical insight into the setup. %João Costa, David Hilditch and José Natário for useful discussions and valuable comments on the manuscript.
FR was supported by FCT/Portugal through the PhD scholarship UI/BD/152068/2021, partially supported by FCT/Portugal through CAMGSD, IST-ID,
projects UIDB/04459/2020 and UIDP/04459/2020; and partially supported by the H2020-MSCA-2022-SE project Einstein--Waves, GA No.101131233.
AVV thanks FCT for financial support through Project~No.~UIDB/00099/2020 and for the FCT funding with DOI 10.54499/DL57/2016/CP1384/CT0090. 
Acknowledged is also the financial support by the EU's H2020 ERC Advanced Grant Gravitas-101052587, and that from the VILLUM Foundation (grant VIL37766) and the DNRF Chair program (grant DNRF162) by the Danish National Research Foundation. This project received funding from EU's Horizon 2020 MSCA grants 101007855 and 101131233.

\appendix

\section{Wave equations} \label{eqsappendix}

Explicit expressions for the wave equations implemented as well as further derivation details are given here.

\subsection{Flat case  with conformal time} \label{eqsappendix_conformal}

To write the wave equation \eref{wave_eqn} in terms of the metric components, we use the common 3+1 ansatz in numerical relativity
\begin{equation}\label{ansatzflat}
ds^2 = - \left(\alpha^2-\gamma_{rr}\,{\beta^{r}}^2\right) d\tau^2 + 2\,\gamma_{rr}\,\beta^{r} d\tau\,dr +  \gamma_{rr}\, dr^2 +  \gamma_{\theta\theta}\, r^2\, d\sigma^2 .
\end{equation} 
The lapse $\alpha$ is not to be confused with the equally-denoted decay parameter that appears in \sref{linearresults}.

The linear wave equation including the time-reduction variable $\Pi=\dot \phi$ and in terms of the metric components of the 3+1 decomposed spacetime metric $g$  above follows.  Dots denote derivatives with respect to the (hyperboloidal) time  and primes indicate derivatives with respect to the (compactified) radial coordinate $r$. 
\begin{subequations}\label{eflatmetric}
\begin{eqnarray}
\dot\phi &=& \Pi, \\
\dot\Pi &=& \frac{\Pi  \dot{\alpha }}{\alpha }-\frac{{\beta^r} \dot{\alpha } \phi '}{\alpha }+\dot{{\beta^r}} \phi
   '+\frac{{\beta^r} \dot{{\gamma_{\theta\theta}}} \phi '}{{\gamma_{\theta\theta}}}-\frac{\Pi  \dot{{\gamma_{\theta\theta}}}}{{\gamma_{\theta\theta}}}+\frac{{\beta^r} \dot{{\gamma_{rr}}} \phi '}{2 {\gamma_{rr}}}-\frac{\Pi \dot{{\gamma_{rr}}}}{2 {\gamma_{rr}}}+\frac{{\beta^r}^2 \alpha ' \phi '}{\alpha }-\frac{{\beta^r} \Pi  \alpha '}{\alpha}-{\beta^r}^2 \phi ''+\Pi  {\beta^r}' + \nonumber \\ && -2 {\beta^r} {\beta^r}' \phi '+2 {\beta^r} \Pi'-\frac{{\beta^r}^2 {\gamma_{\theta\theta}}' \phi '}{{\gamma_{\theta\theta}}}+\frac{{\beta^r} \Pi  {\gamma_{\theta\theta}}'}{{\gamma_{\theta\theta}}}-\frac{{\beta^r}^2 {\gamma_{rr}}' \phi '}{2 {\gamma_{rr}}}+\frac{{\beta^r} \Pi  {\gamma_{rr}}'}{2 {\gamma_{rr}}}-\frac{\alpha ^2 {\gamma_{rr}}' \phi '}{2 {\gamma_{rr}}^2}+\frac{\alpha ^2 \phi ''}{{\gamma_{rr}}}+\frac{\alpha  \alpha ' \phi'}{{\gamma_{rr}}}+ \nonumber \\ && + \frac{\alpha ^2 {\gamma_{\theta\theta}}' \phi '}{{\gamma_{\theta\theta}} {\gamma_{rr}}} +\frac{2 \alpha ^2 \phi'}{{\gamma_{rr}} r}-\frac{2 {\beta^r}^2 \phi '}{r}+\frac{2 {\beta^r} \Pi }{r} . 
\end{eqnarray}
\end{subequations}

To obtain the expressions of the metric components, we compare \eref{fsthyp} to the spherically symmetric ansatz \eref{ansatzflat}, obtaining
%, to appear in the wave equation and to be substituted in by the corresponding expressions from \eref{fsthyp}

%Their background values are thus
%{Therefore, the values of $\alpha$, $\gamma_{rr}$, $\beta^r$ and $\gamma_{\theta \theta}$ are:
\begin{subequations}\label{metriccompflat}
\begin{eqnarray}
&&\gamma_{rr}= -\frac{a^2 \left(\aconf^4 \left(h'\right)^2-\aconf^2-r^2
   \left(\aconf'\right)^2+2 \aconf r \aconf'\right)}{\aconf^4},
\quad \gamma_{\theta\theta}= \frac{a^2}{\aconf^2}, \\
&&\alpha = a\sqrt{\frac{ \left(\aconf-r \aconf'\right)^2}{-\aconf^4   \left(h'\right)^2+\aconf^2+r^2 \left(\aconf'\right)^2-2 \aconf r \aconf'}},
\quad\beta^r = -\frac{\aconf^4 h'}{-\aconf^4 \left(h'\right)^2+\aconf^2+r^2 \left(\aconf'\right)^2-2 \aconf r   \aconf'},
\end{eqnarray}
\end{subequations}
with $a=a(\tau+h)$ using \eref{scalefacflat}, $h=h(r)$ given by \eref{hfuncflat} and $\aconf=\aconf(r)$ by \eref{confflat}. 

After substitution of the above expressions for the metric components into \eref{eflatmetric}, the final expression for the wave equation is
\begin{subequations}\label{eflat}
\begin{eqnarray}
\dot\phi &=& \Pi, \\
% without the extra -aa in the height function: \dot\Pi &=& \frac{\Pi  \left(2 r^2 \left(\sqrt{\aaasquared+\frac{r^2}{\aconf ^2}}-p t+t\right)+\aaasquared \aconf ^2 \left((p-3)    \left(-\sqrt{\aaasquared+\frac{r^2}{\aconf ^2}}\right)-3 (p-1) t\right)\right)}{\aaasquared (p-1) \aconf ^2    \sqrt{\aaasquared+\frac{r^2}{\aconf ^2}} \left(\sqrt{\aaasquared+\frac{r^2}{\aconf ^2}}+t\right)} + \nonumber \\ && -\frac{2 r \aconf  \Pi '    \sqrt{\aaasquared+\frac{r^2}{\aconf ^2}}}{\aaasquared \left(\aconf -r \aconf '\right)} +\frac{\aconf ^2 \phi '' \left(\aaasquared \aconf    ^2+r^2\right)}{\aaasquared \left(\aconf -r \aconf '\right)^2}-\frac{\aconf ^2 \phi ' \left(\aaasquared \aconf ^2+r^2\right) \left(r \left(r    \aconf ''-2 \aconf '\right)+2 \aconf \right)}{\aaasquared r \left(r \aconf '-\aconf \right)^3}
\dot\Pi &=& \frac{\Pi  \left(2 r^2 \left(\sqrt{\aaasquared+\frac{r^2}{\aconf ^2}}+\aaa (p-1)-p t+t\right)+\aaasquared \aconf ^2 \left((p-3)  \left(-\sqrt{\aaasquared+\frac{r^2}{\aconf ^2}}\right)+3 \aaa (p-1)-3 (p-1) t\right)\right)}{\aaasquared (p-1) \aconf ^2  \sqrt{\aaasquared+\frac{r^2}{\aconf ^2}} \left(\sqrt{\aaasquared+\frac{r^2}{\aconf ^2}}-\aaa+t\right)}+ \nonumber \\ &&-\frac{2 r \aconf  \Pi '  \sqrt{\aaasquared+\frac{r^2}{\aconf ^2}}}{\aaasquared \left(\aconf -r \aconf '\right)}+\frac{\aconf ^2 \phi '' \left(\aaasquared \aconf  ^2+r^2\right)}{\aaasquared \left(\aconf -r \aconf '\right)^2}-\frac{\aconf ^2 \phi ' \left(\aaasquared \aconf ^2+r^2\right) \left(r   \left(r \aconf ''-2 \aconf '\right)+2 \aconf \right)}{\aaasquared r \left(r \aconf '-\aconf\right)^3} , 
\end{eqnarray}
\end{subequations}
where $\Omega$ is given by \eqref{confflat}.
%{\avv expression above including the $-aa$ term in $h$}

\subsection{ Flat case with physical time and time-dependent height function}

Given the 3+1 ansatz (with respect to physical time):
\begin{equation}\label{ansatzflatphysical}
ds^2 = - \left(\alpha^2-\gamma_{rr}\,{\beta^{r}}^2\right) dt^2 + 2\,\gamma_{rr}\,\beta^{r} dt\,dr +  \gamma_{rr}\, dr^2 +  \gamma_{\theta\theta}\, r^2\, d\sigma^2 ,
\end{equation}
we follow the procedure used for \eref{metriccompflat} above to get
\begin{subequations}
\begin{eqnarray}\label{metriccompflattimedep}
&&\gamma_{rr}= \frac{a^2 \left(\aconf-r \aconf'\right)^2-\aconf^4 \left(h'\right)^2}{\aconf^4},
\quad \gamma_{\theta\theta}= \frac{a^2}{\aconf^2}, \\
&&\alpha = \sqrt{\left(\dot h+1\right)^2 \left(\frac{\aconf^4 \left(h'\right)^2}{a^2 \left(\aconf-r
   \aconf'\right)^2-\aconf^4 \left(h'\right)^2}+1\right)},
\quad\beta^r = \frac{\aconf^4 h'
   \left(\dot h+1\right)}{\aconf^4 \left(h'\right)^2-a^2 \left(\aconf-r \aconf'\right)^2},
\end{eqnarray}
\end{subequations}
where dots denote differentiation with respect to the hyperboloidal time $t$ defined in \eqref{ttrafo}.
Substituting the scale factor, the height function with \eref{tdepheightfunc}, and the compactification factor with \eref{confflat} yields
\begin{subequations}
\begin{eqnarray}
&&\gamma_{rr}= \gamma_{\theta\theta}= \frac{4 \aaasquared \left((p-1) \left(\frac{2 \aaa}{r^2-1}-\C(t)\right)\right)^{-\frac{2 p}{p-1}}}{\left(r^2-1\right)^2}, \\
&&\alpha = -\frac{\left(r^2+1\right) \left| \dot\C(t)\right| \left((p-1) \left(\frac{2 \aaa}{r^2-1}-\C(t)\right)\right)^{-\frac{p}{p-1}}}{r^2-1},
\quad\beta^r = -\frac{r \dot\C(t)}{\aaa}.
\end{eqnarray}
\end{subequations}

The expressions for the linear wave equations corresponding to each of the choices of $C(t)$ mentioned in the main text follow:
\begin{enumerate}
\item $\C(t)=t^{1-p}$:
\begin{subequations}\label{eflavio1}
\begin{eqnarray}
\dot\phi &=& \Pi, \\
\dot\Pi &=& \frac{\Pi  t^{-p-1} \left(-2 \aaasquared p \left(r^4-1\right) t^{2 p}+\aaa \left(3 r^4+2 r^2+3\right) \left(p    \left(r^2-1\right)+2\right) t^{p+1}+(p-1) \left(3 r^6-r^4+r^2-3\right) t^2\right)}{\aaa \left(r^2-1\right) \left(r^2+1\right) \left(2    \aaa t^p-r^2 t+t\right)} + \nonumber \\ && + \frac{(p-1)^2 \left(r^2-1\right)^2 t^{-2 p} \phi ''}{4 \aaasquared}+\frac{(p-1)^2 \left(r^2-1\right)^2 t^{-2 p} \phi '}{2 \aaasquared    \left(r^3+r\right)} +\frac{2 (p-1) r t^{-p} \Pi '}{\aaa} 
\end{eqnarray}
\end{subequations}

\item $\C(t)=t$:
\begin{subequations}\label{eflavio2}
\begin{eqnarray}
\dot\phi &=& \Pi, \\
\dot\Pi &=& \frac{\left(r^2-1\right)^2 \phi ''}{4 \aaasquared}+\frac{\left(r^2-1\right)^2 \phi '}{2 \aaasquared \left(r^3+r\right)}-\frac{\Pi  \left(2    \aaa \left(p \left(r^6+r^2-2\right)+3 r^4+2 r^2+3\right)+(p-1) \left(3 r^6-r^4+r^2-3\right) t\right)}{\aaa (1-p) \left(r^2+1\right) \left(1-r^2\right) \left(2 \aaa+t(1-r^2)\right)} + \nonumber \\ &&-\frac{2 r \Pi '}{\aaa}
\end{eqnarray}
\end{subequations}

\item $\C(t)=-1/t$:
\begin{subequations}\label{eflavio3}
\begin{eqnarray}
\dot\phi &=& \Pi, \\
\dot\Pi &=& 
\frac{\Pi}{\aaa (1-p) \left(r^2+1\right) \left(1-r^2\right) t^2 \left(2 \aaa t+r^2-1\right)}  
\left( r^4 \left(4 \aaasquared t^2-8 \aaa t+1\right)-4 \aaasquared t^2+r^6 (2 \aaa t-3)-r^2 (6 \aaa t+1) \right. + \nonumber \\ && \left. + p \left(r^4 \left(-4 \aaasquared t^2+2 \aaa t-1\right)+4 \aaasquared t^2+r^6 (3-4 \aaa t)+2 \aaa
   t+r^2-3\right) -4\aaa t+3\right)
 +\frac{\left(r^2-1\right)^2 \phi ''}{4 \aaasquared t^4} + \nonumber \\ && +\frac{\left(r^2-1\right)^2 \phi '}{2 \aaasquared \left(r^3+r\right) t^4}-\frac{2 r \Pi '}{\aaa t^2}
\end{eqnarray}
\end{subequations}

\end{enumerate}

\subsection{Hyperbolic case with conformal time}

Given the hyperbolic spherically symmetric ansatz
\begin{equation}\label{ansatzhyperb}
ds^2 = - \left(\alpha^2-\gamma_{\rho\rho}\,{\beta^{\rho}}^2\right) d\tau^2 + 2\,\gamma_{\rho\rho}\,\beta^{\rho} d\tau\,d\rho +  \gamma_{\rho\rho}\, d\rho^2 +  \gamma_{\theta\theta}\, \sinh^2\rho\, d\sigma^2,
\end{equation} 
we can retrieve the expression for the linear wave equation as follows. As before, dots denote derivative with respect to the (hyperboloidal) conformal time $\tau$ and primes indicate derivatives with respect to the compactified coordinate $\rho$:
\begin{subequations}\label{ehyperbmetric}
\begin{eqnarray}
\dot\phi &=& \Pi, \\
\dot\Pi &=& \frac{\Pi  \dot{\alpha }}{\alpha }-\frac{{\beta^\rho} \dot{\alpha } \phi '}{\alpha }+\dot{{\beta^\rho}} \phi'+\frac{{\beta^\rho} \dot{{\gamma_{\theta\theta}}} \phi '}{{\gamma_{\theta\theta}}}-\frac{\Pi  \dot{{\gamma_{\theta\theta}}}}{{\gamma_{\theta\theta}}}+\frac{{\beta^\rho} \dot{{\gamma_{\rho\rho}}} \phi '}{2 {\gamma_{\rho\rho}}}-\frac{\Pi \dot{{\gamma_{\rho\rho}}}}{2 {\gamma_{\rho\rho}}}+\frac{{\beta^\rho}^2 \alpha ' \phi '}{\alpha }-\frac{{\beta^\rho} \Pi \alpha '}{\alpha }-{\beta^\rho}^2 \phi ''+\Pi  {\beta^\rho}'+ \nonumber \\ && -2 {\beta^\rho} {\beta^\rho}' \phi '+2 {\beta^\rho} \Pi'-2 {\beta^\rho}^2 \coth\rho\ \phi '+2 {\beta^\rho} \coth\rho\ \Pi +\frac{2 \alpha ^2 \coth\rho\ \phi '}{{\gamma_{\rho\rho}}} -\frac{{\beta^\rho}^2 {\gamma_{\theta\theta}}' \phi '}{{\gamma_{\theta\theta}}}+\frac{{\beta^\rho} \Pi  {\gamma_{\theta\theta}}'}{{\gamma_{\theta\theta}}} + \nonumber \\ && +\frac{\alpha ^2 {\gamma_{\theta\theta}}' \phi '}{{\gamma_{\theta\theta}} {\gamma_{\rho\rho}}}-\frac{{\beta^\rho}^2 {\gamma_{\rho\rho}}' \phi '}{2 {\gamma_{\rho\rho}}}+\frac{{\beta^\rho} \Pi  {\gamma_{\rho\rho}}'}{2 {\gamma_{\rho\rho}}}-\frac{\alpha ^2 {\gamma_{\rho\rho}}' \phi '}{2 {\gamma_{\rho\rho}}^2}+\frac{\alpha ^2 \phi ''}{{\gamma_{\rho\rho}}}+\frac{\alpha  \alpha ' \phi '}{{\gamma_{\rho\rho}}} . 
\end{eqnarray}
\end{subequations}

By comparing \eref{hyperhyp} with \eqref{ansatzhyperb}, we obtain for the metric components:
\begin{subequations}\label{metriccomphyperb}
\begin{eqnarray}
&&\gamma_{\rho\rho} = -\frac{a^2 \left(\aconf^4 \left(h'\right)^2+\aconf^2 \left(\sinh^2\rho \left(h'\right)^2-\cosh^2\rho\right)+\aconf \aconf' \sinh(2 \rho )-\sinh^2\rho \left(\aconf'\right)^2\right)}{\aconf^2 \left(\aconf^2+\sinh^2\rho\right)}, \\
&&\gamma_{\theta \theta} = \frac{a^2}{\aconf^2},
\quad \alpha = a\sqrt{-\frac{\left(\aconf \cosh\rho-\aconf'\sinh\rho\right)^2}{\aconf^4 \left(h'\right)^2+\aconf^2 \left(\sinh^2\rho   \left(h'\right)^2-\cosh^2\rho\right)+\aconf \aconf' \sinh (2 \rho )-\sinh^2\rho \left(\aconf'\right)^2}}, \\
&&\beta^\rho = \frac{\aconf^2 \left(\aconf^2+\sinh^2\rho\right) h'}{\aconf^4 \left(h'\right)^2+\aconf^2   \left(\sinh^2\rho \left(h'\right)^2-\cosh^2\rho\right)+\aconf \aconf' \sinh (2 \rho )-\sinh^2\rho    \left(\aconf'\right)^2} ,
\end{eqnarray}
\end{subequations}
where $a=a(\tau+h)$ using \eref{scalefachyperb}, $h=h(\rho)$ as set by \eref{hfunchyperb} and $\aconf=\aconf(\rho)$ by \eref{confhyperb}. 

The final evolution equations, using the shorthand $\shorthand=\sqrt{\aaasquared+\left[\sinh ^{-1}\left(\frac{\sinh (\rho )}{\aconf}\right)\right]^2}$, are
\begin{subequations}\label{ehyperb}
\begin{eqnarray}
\dot\phi &=& \Pi, \\
\dot\Pi &=& \frac{\shorthand^2 \aconf ^2 \phi '' \left(\sinh ^2(\rho )+\aconf ^2\right)}{\aaasquared \left(\aconf  \cosh (\rho )-\sinh (\rho )
   \aconf '\right)^2}
   + \nonumber \\ && 
   +\frac{\shorthand^2 \aconf  \phi ' }{2 \aaasquared \left(\aconf  \coth (\rho )-\aconf'\right)^3}
   \left[\aconf ^4 (\cosh (2 \rho )+3) \text{csch}^4(\rho )+2 \aconf ^2 (\cosh
   (2 \rho )+2) \text{csch}^2(\rho )
   + \right. \nonumber \\ && \left. 
   +\aconf  \left(2 \aconf ''-8 \coth (\rho ) \aconf '\right)-2 \aconf ^3 \text{csch}^2(\rho ) \left(2
   \coth (\rho ) \aconf '-\aconf ''\right)+2 \left(\aconf '\right)^2\right]
   + \nonumber \\ && 
   +\frac{\Pi}{\aaasquared \sqrt{\shorthand^2} \aconf  \sqrt{\frac{\sinh ^2(\rho
   )}{\aconf ^2}+1} \left(\aconf  \cosh (\rho )-\sinh (\rho ) \aconf '\right)^2}
   \left[\aconf ^2 \left\{\sinh (2 \rho ) \aconf ' \sqrt{\frac{\sinh ^2(\rho )}{\aconf ^2}+1} 
   \cdot
   \right.\right. \nonumber \\ && \left.\left. 
   \left(2 \aaasquared
   \shorthand \coth \left(\frac{1}{2} (3 w+1) (-\aaa+\shorthand+t)\right)+\aaasquared+2 \shorthand \sinh
   ^{-1}\left(\frac{\sinh (\rho )}{\aconf }\right)^2 \coth \left(\frac{1}{2} (3 w+1) (-\aaa+\shorthand+t)\right)\right)
   +\right.\right. \nonumber \\ && \left.\left. 
   -2 \shorthand^2 \sinh (\rho ) \left(\aconf '\right)^2 \sinh ^{-1}\left(\frac{\sinh (\rho )}{\aconf }\right)-2 \shorthand^2
   \sinh (\rho ) \cosh ^2(\rho ) \sinh ^{-1}\left(\frac{\sinh (\rho )}{\aconf }\right)\right\}
   + \right. \nonumber \\ && \left. 
   -\aconf  \sinh ^2(\rho ) \aconf ' \left\{\aconf
   ' \sqrt{\frac{\sinh ^2(\rho )}{\aconf ^2}+1} \left(2 \aaasquared \shorthand \coth \left(\frac{1}{2} (3 w+1)
   (-\aaa+\shorthand+t)\right)+\aaasquared
   +\right.\right.\right. \nonumber \\ && \left.\left.\left.
   +2 \shorthand \sinh ^{-1}\left(\frac{\sinh (\rho )}{\aconf }\right)^2 \coth
   \left(\frac{1}{2} (3 w+1) (-\aaa+\shorthand+t)\right)\right)-4 \shorthand^2 \cosh (\rho ) \sinh ^{-1}\left(\frac{\sinh
   (\rho )}{\aconf }\right)\right\}
   + \right. \nonumber \\ && \left. 
   +\aconf ^3 \left\{4 \shorthand^2 \cosh (\rho ) \aconf ' \sinh ^{-1}\left(\frac{\sinh (\rho )}{\aconf
   }\right)-\cosh ^2(\rho ) \sqrt{\frac{\sinh ^2(\rho )}{\aconf ^2}+1} 
   \cdot
   \right.\right. \nonumber \\ && \left.\left.
   \left(2 \aaasquared \shorthand \coth \left(\frac{1}{2} (3 w+1)
   (-\aaa+\shorthand+t)\right)+\aaasquared+2 \shorthand \sinh ^{-1}\left(\frac{\sinh (\rho )}{\aconf }\right)^2 \coth
   \left(\frac{1}{2} (3 w+1) (-\aaa+\shorthand+t)\right)\right)\right\}
   + \right. \nonumber \\ && \left.
   -2 \shorthand^2 \aconf ^4 \cosh (\rho ) \coth (\rho
   ) \sinh ^{-1}\left(\frac{\sinh (\rho )}{\aconf }\right)-2 \shorthand^2 \sinh ^3(\rho ) \left(\aconf '\right)^2 \sinh
   ^{-1}\left(\frac{\sinh (\rho )}{\aconf }\right)\right]
   + \nonumber \\ && 
   -\frac{2 \shorthand \Pi ' \sinh
   ^{-1}\left(\frac{\sinh (\rho )}{\aconf }\right) \left(\sinh ^2(\rho )+\aconf ^2\right)}{\aaasquared \sqrt{\frac{\sinh ^2(\rho )}{\aconf
   ^2}+1} \left(\aconf  \cosh (\rho )-\sinh (\rho ) \aconf '\right)} . 
%+ \nonumber \\ &&   
\end{eqnarray}
\end{subequations}
%{\avv @Flavio: do you prefer $\sinh ^{-1}$ or $\arcsinh$?}{\fr @Alex: I don't have strong opinions, the current choice looks good to me!}

\section{Pseudo-codes} \label{appendixAlgorithm}

In the following, we describe an algorithm that takes as input the data obtained from the numerical evolution with respect to a hyperboloidal foliation, and provides data with respect to a $\tilde T$-constant foliation, where $\tilde T$ can be any unrescaled time function. In our implementation, we compared hyperboloidal slices with those slices determined by the conformal time coordinate, for the sake of simplicity. Few algebraic steps easily allow to further convert the results in terms of the physical, unrescaled, time coordinate.
In the pseudo-code, we store the hyperboloidal times in variables containing the prefix \textit{hyp}, whereas the unrescaled times are not marked by any prefix.

In addition to the functions specified in the pseudo-code, we also use the following mathematical functions and notations: 
\begin{itemize}
\item \texttt{Solve(eqn, x)}, which provides the solutions of the equation \texttt{eqn} in terms of the unknown \texttt{x},
\item \texttt{Interpolate(data\_set)}, which provides a function that interpolates the given points. This can be thought of as the analogue of the  \texttt{Interpolation} built--in symbol in Mathematica. The result of the function \texttt{Interpolate} is a mathematical function that can be evaluated at a given point in the allowed range,
\item \texttt{Length}, which gives the length of a list,
\item \texttt{Abs}, which is the absolute value function,
\item The notation $\text{Elements}_i$, which denotes the $i$-th component of the list Elements.
\end{itemize}

The algorithm proceeds as follows. First of all, we need to find the points of intersection between the hyperboloidal slices at our disposal and the $\tilde T$-constant slices that we later consider, see Fig. \ref{recovery_diagram}. Since we only know the values of the solution on a discrete subset of each hyperboloidal slice, we interpolate the solution on each slice so to compute the points of intersection with higher accuracy. The hyperboloidal data obtained from the evolution consist of a list of lists (see already the input data in the full algorithm at the end of this appendix), so it is sufficient to store each interpolated function corresponding to a hyperboloidal slice as the element of a list:
\begin{algorithmic}
 \ForAll{$i = 0, 1, \ldots, \texttt{Length}(\text{hyp\_Data})-1$}
	 \State $\text{interpolated\_solution}_i \gets \texttt{Interpolate}( (\text{hyp\_Data}_i)_1)$
	 \EndFor
\end{algorithmic} 
Now, for a given point $\tilde r_0$ in the grid at the initial hyperboloidal time $T_0$, it is possible to find the intersecting $\tilde T$-slice by the relation
\[
  \tilde T - h(\tilde r) = T,
\]
$h$ being the height function of the hyperboloidal foliation. The above can be encoded in the following pseudo-code: 
\begin{algorithmic}
\State $t \gets \texttt{Solve}(x - \text{height\_fct}(r) = (\text{hyp\_Data}_0)_0, x)$
\end{algorithmic}
Now that we chose a $\tilde T$-constant slice, we can find the values $\tilde{r}_1$, $\tilde{r}_2, \ldots$ corresponding to the intersections of the $\tilde T$-constant slice with the second hyperboloidal slice, third hyperboloidal slice, ..., respectively, and save them in a new list of lists:
\begin{algorithmic}
\State $m \gets 0$
\State \ForAll{$\text{hyp\_time} = 1, 2, \ldots, \texttt{Length}(\text{hyp\_Data}) - 1$}
	\State \State $\text{m\_th\_intersection} \gets \texttt{Solve}(t - \text{height\_fct}(x) = (\text{hyp\_Data}_{\text{hyp\_time}})_0, x)$
 \State\State $((\text{t\_Data}_0)_1)_m \gets \texttt{Abs}(\text{interpolated\_solution}_m(\text{m\_th\_intersection}))$
 \State \State $m \gets m + 1$
	\State \EndFor
\end{algorithmic}
We stored the absolute value of the above numbers since we are ultimately interested in the supremum norm of the solution. The procedure can then be iterated for each $\tilde T$-constant slice, by repeating the above starting from values of $\tilde r$ larger than $\tilde r_0$.

The full algorithm is the following.
\begin{algorithmic}
\Function{Convert\_Foliation}{hyp\_Data, grid\_length, height\_fct}
	\State \textbf{in:} list of lists hyp\_Data $ = [ [ $ hyp\_t1, hyp\_slice1 $ ], [ $ hyp\_t2, hyp\_slice2 $ ], \ldots]$,
	\State where hyp\_slice1 $= [[ $ sol\_x1, sol\_y1 $], [ $ sol\_x2, sol\_y2 $], \ldots]$;
	\State integer grid\_length; 
	\State function height\_fct defining the hyperboloidal time coordinate.
	\State \textbf{out: } list of lists t\_Data, having a similar expression to hyp\_data, containing the values of the solution on each $t$-constant slice.
	\newline
	\State \ForAll{$i = 0, 1, \ldots, \texttt{Length}(\text{hyp\_Data})-1$}
	\State \State $\text{interpolated\_solution}_i \gets \texttt{Interpolate}( (\text{hyp\_Data}_i)_1)$
	\State \EndFor
	\State $l \gets 0$
	\State \ForAll{$r=0, 1, \ldots, \text{grid\_length}-1$}
	\State \State $t \gets \texttt{Solve}(x - \text{height\_fct}(r) = (\text{hyp\_Data}_0)_0, x)$
	\State \State $(\text{t\_Data}_l)_0 \gets t$
	\State \State $m \gets 0$
	\State \State \ForAll{$\text{hyp\_time} =  1, 2, \ldots, \texttt{Length}(\text{hyp\_Data}) - 1$}
	\State \State \State $\text{m\_th\_intersection} \gets \texttt{Solve}(t - \text{height\_fct}(x) = (\text{hyp\_Data}_{\text{hyp\_time}})_0, x)$
	\State \State \State $((\text{t\_Data}_l)_1)_m \gets \texttt{Abs}(\text{interpolated\_solution}_m(\text{m\_th\_intersection}))$
	\State \State \State $m \gets m + 1$
	\State \State \EndFor
	\State $l \gets l+1$
	\State \EndFor
	\State \Return \text{t\_Data}
\EndFunction
\end{algorithmic}
The above is the main structure of the algorithm we used. In the Mathematica code we adopted, further refinements were implemented. For instance, the case in which no intersection between slices is found is taken into account as well. Starting from later hyperboloidal slices or larger values of the radial coordinate was also implemented, so to investigate subregions of the spacetime and save computation time.

%%%%%%%%%%%%%%%%%%%%%%%%%%%%%%%%%%%%%%%%%%%%%%%%%%%%%%%%%%%%%%%%%%%%%%%%%%%%%%%%%%%%%%%%%%%%%%%%%%%%%%%%%%%%%%%%%%%%%%%%%%%%%%%%%%%%%%
%\section*{References}
%\bibliographystyle{unsrt}
%\addcontentsline{toc}{chapter}{References}
\bibliography{refs}

\end{document}